\documentclass{article}

\usepackage[final,nonatbib]{neurips_2021}

\usepackage[utf8]{inputenc} 
\usepackage[T1]{fontenc}    
\usepackage{hyperref}       
\usepackage{url}            
\usepackage{booktabs}       
\usepackage{amsfonts}       
\usepackage{nicefrac}       
\usepackage{microtype}      
\usepackage{xcolor}         

\usepackage{placeins}
\usepackage{algorithmic}

\usepackage{amsmath,amssymb,amsfonts}
\usepackage{bbm}
\usepackage{graphicx}
\usepackage{graphics}
\usepackage{multirow,makecell} 
\usepackage{booktabs}
\usepackage{amssymb} 
\usepackage{wrapfig}
\usepackage{hhline}
\usepackage{appendix}
\usepackage{caption}
\usepackage{subcaption}
\usepackage{csquotes}
\usepackage{adjustbox}
\usepackage{lettrine}

\usepackage{dblfloatfix}
\newcommand{\norm}[1]{\left\lVert#1\right\rVert}

\title{CROCS: Clustering and Retrieval of Cardiac Signals Based on Patient Disease Class, Sex, and Age}

\author{%
  Dani Kiyasseh, Tingting Zhu, \& David Clifton \\
  Department of Engineering Science\\
  University of Oxford\\
  Oxford, UK \\
  \texttt{\{dani.kiyasseh,tingting.zhu,david.clifton\}@eng.ox.ac.uk}  \\
}

\begin{document}

\maketitle

\begin{abstract}

The process of manually searching for relevant instances in, and extracting information from, clinical databases underpin a multitude of clinical tasks. Such tasks include disease diagnosis, clinical trial recruitment, and continuing medical education. This manual search-and-extract process, however, has been hampered by the growth of large-scale clinical databases and the increased prevalence of unlabelled instances. To address this challenge, we propose a supervised contrastive learning framework, CROCS, where representations of cardiac signals associated with a set of patient-specific attributes (e.g., disease class, sex, age) are attracted to learnable embeddings entitled clinical prototypes. We exploit such prototypes for \textit{both} the clustering and retrieval of unlabelled cardiac signals based on multiple patient attributes. We show that CROCS outperforms the state-of-the-art method, DTC, when clustering and also retrieves relevant cardiac signals from a large database. We also show that clinical prototypes adopt a semantically meaningful arrangement based on patient attributes and thus confer a high degree of interpretability. 

\end{abstract}

\section{Introduction}

Clinical databases can comprise instances that are either unlabelled or labelled with patient attribute information, such as disease class, sex, and age. The process of manually searching for relevant instances in, and extracting information from, such databases underpin a multitude of tasks \cite{Shivade2014}. For example, clinicians extract a disease diagnosis from patient data, researchers involved in clinical trials search for and recruit patients satisfying specific inclusion criteria \cite{Murthy2004}, and educators retrieve relevant information as part of the continuing medical education scheme \cite{Pourmand2015}. This manual search-and-extract process, however, is hampered by the rapid growth of large-scale clinical databases and the increased prevalence of unlabelled instances; those for which patient attribute information is unavailable.

Given such a setting, in this paper, we are interested in addressing two questions: given a large, unlabelled clinical database, (1) \textit{how do we extract attribute information from such unlabelled instances?} and (2) \textit{how do we reliably search for and retrieve relevant instances?} To address the former, the task of clustering holds value. In this setting, a centroid groups together instances that share some similarities. Recent research has focused on exploiting existing clustering algorithms, such as $k$-means, to group similar patients from electronic heath record (EHR) data \cite{Huang2019,Landi2020}. Such methods, however, are exclusively unsupervised; they do not exploit patient attribute information. To address the second question, the task of information retrieval holds promise. In this setting, a query associated with a set of desired attributes is exploited to retrieve a relevant instance. Recent research has focused predominantly on retrieving medical images \cite{Saritha2019}, clinical text \cite{Wang2017}, and EHR data \cite{Chamberlin2019}, with minimal emphasis on medical time-series data \cite{Goodwin2016}. These methods do not extend to cardiac time-series data nor do they account for search based on multiple patient attributes. Most notably, previous work performs either clustering or retrieval, and not both.

In this work, we address both questions while exploiting large-scale electrocardiogram (ECG) databases comprising patient attribute information. Our contributions are the following: (1) we propose a supervised contrastive learning framework, CROCS, in which we attract representations of cardiac signals associated with a unique set of patient attributes to embeddings, entitled clinical prototypes. Such attribute-specific prototypes, which create \enquote{islands} of similar representations \cite{Hinton2021}, allow for \textit{both} the clustering and retrieval of cardiac signals based on \textit{multiple} patient attributes. (2) We show that CROCS outperforms the state-of-the-art method, DTC, in the clustering setting and retrieves relevant cardiac signals from a large database. At the same time, clinical prototypes adopt a semantically meaningful arrangement and thus confer a high degree of interpretability. 


\section{Related work}

\paragraph{Clinical representation learning and clustering} Learning meaningful representations of clinical data is an ongoing research endeavour. Recent research has focused on learning representations from EHR data \cite{Gee2019,Liu2019,Li2020,Biswal2020,Darabi2020} and via auto-encoders, which are then clustered using existing methods, such as $k$-means \cite{Huang2019,Landi2020}. As for time-series data, auto-encoders are learned with \cite{Ma2019} or without \cite{Madiraju2018} an auxiliary clustering objective, salient features (shapelets) are identified in an unsupervised manner \cite{Grabocka2014,Zhang2018}, and patient-specific representations are learned via contrastive learning \cite{Kiyasseh2020CLOCS}. Li \textit{et al.} \cite{Li2020PCL} learn prototypes, or representative embeddings, via the ProtoNCE loss and cluster instances using $k$-means. Their work builds upon recent research in the contrastive learning literature \cite{Chen2020,Grill2020,Zhang2020}. Similar to our work is that of Gee \textit{et al.} \cite{Gee2019} where prototypes are learned for the clustering of time-series signals. Their prototypes, however, cannot cluster instances based on multiple patient attributes and do not extend to the retrieval setting. 

\paragraph{Clinical information retrieval (IR)} Retrieving clinical data from a large database has been a longstanding goal of researchers within healthcare \cite{Hersh1990}. Such research has involved the retrieval of clinical documents \cite{Gurulingappa2016,Wang2017,Rhine2017,Wallace2016} where, for example, Avolio \textit{et al.} \cite{Avolio2010} map text queries to an ontology known as SNOMED, before retrieving relevant clinical documents. Recent research has focused on the retrieval of biomedical images \cite{Saritha2019,Chittajallu2019}, and EHR data \cite{Goodwin2018,Wang2019} to discover patient cohorts in a clinical database \cite{Chamberlin2019}. Goodwin \textit{et al.} \cite{Goodwin2016} implement an unsupervised patient cohort retrieval system by exploiting clinical text and time-series data. These approaches, however, do not explore cardiac signals, cannot account for multiple patient attributes, and are unable to also cluster instances. To the best of our knowledge, we are the first to design a learning framework that allows for \textit{both} the clustering and retrieval of cardiac signals based on multiple patient attributes. 

\section{Background}

\paragraph{Supervised clustering} We learn a function, $f_{\theta}: \boldsymbol{x} \in \mathbb{R}^{D} \xrightarrow{} \boldsymbol{v} \in \mathbb{R}^{E}$, parameterized by $\boldsymbol{\theta}$, that maps a $D$-dimensional input, $\boldsymbol{x}$, to an $E$-dimensional representation, $\boldsymbol{v}$. We also have a labelled dataset, $\mathcal{D}_{l} = \{\boldsymbol{x_{i}},A_{i}\}_{i=1}^{N_{l}}$, where each instance, $\boldsymbol{x_{i}}$, is associated with a set of discrete patient attributes, $A_{i}=\{ \alpha_{c}^{i},\alpha_{s}^{i},\alpha_{a}^{i} \}$ where $\alpha_{c}=\mathrm{disease \ class}$, $\alpha_{s}=\mathrm{sex}$ and $\alpha_{a}=\mathrm{age}$. Supervised clustering can involve learning $M \ll N_{l}$ centroids with each representing a unique set of attributes, $\{ \alpha_{c}^{j},\alpha_{s}^{j},\alpha_{a}^{j} \} \in \{ A_{j} \}_{j=1}^{M}$, and grouping similar instances together. Given unlabelled instances, $ \{ \boldsymbol{x_{u}} \}_{u=1}^{N}$, the centroid closest to each representation, $\boldsymbol{v_{u}} = f_{\theta}(\boldsymbol{x_{u}})$, is used to infer the latter's attributes. In this work, we learn cluster centroids which are more formally introduced in Sec.~\ref{sec:prototypes}. 

\paragraph{Information retrieval} IR involves searching through a large, unlabelled dataset, $ \{ \boldsymbol{x_{u}} \}_{u=1}^{N}$, and retrieving a relevant instance, $\boldsymbol{x_{u}}$. However, relevance, defined based on whether an instance satisfies some criteria, is difficult to ascertain when instances are \textit{unlabelled}. Typically, a query in the form of an embedding which represents a desired set of attributes, $A_{j}$, retrieves its closest (and most relevant) representation, $\boldsymbol{v_{u}} = f_{\theta}(\boldsymbol{x_{u}})$, and infers the latter's attributes. In this work, we learn a set of query embeddings. As will become apparent in Sec.~\ref{sec:design}, these embeddings can also be treated as centroids, like those outlined in supervised clustering, and will thus serve a dual purpose. 

\section{Methods}
\label{sec:methods}

\subsection{Attribute-specific clinical prototypes}
\label{sec:prototypes}
\begin{figure}[!h]
    \centering
    \begin{subfigure}{0.9\textwidth}
    \includegraphics[width=1\textwidth]{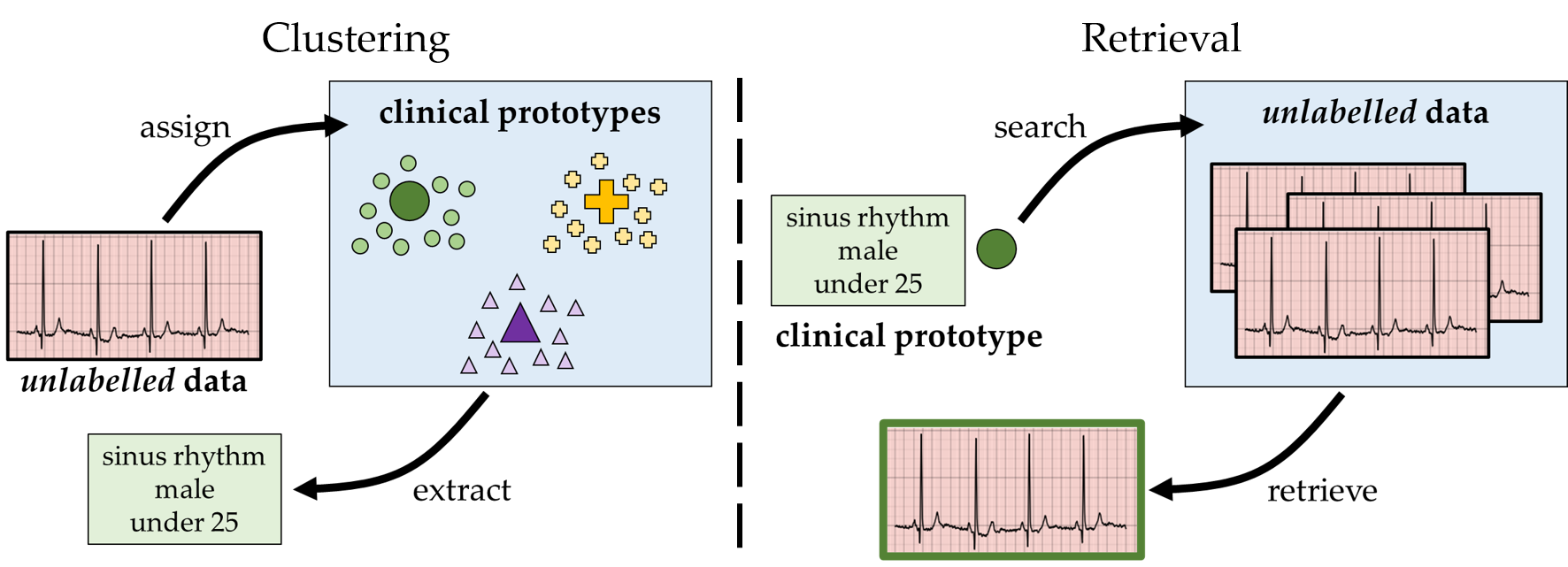}
    \end{subfigure}
    \caption{\textbf{Clinical prototypes are exploited for attribute-specific clustering and retrieval of cardiac signals.} For clustering, we exploit prototypes as centroids of clusters to which \textit{unlabelled} instances are assigned. Such an assignment is associated with a set of attributes, such as disease class, sex, and age. For retrieval, we exploit each prototype as a query, associated with a set of attributes, to search through an \textit{unlabelled} database and retrieve instances to which it is most similar.}
    \label{fig:retrieval_and_clustering}
\end{figure}

We aim to learn embeddings, referred to as clinical prototypes, that can be exploited for \textit{both} the clustering and retrieval of cardiac signals based on multiple patient attributes. In the clustering setting, the goal is to annotate \textit{unlabelled} instances with a set of patient attributes. To that end, we exploit clinical prototypes as centroids of clusters to which such unlabelled instances are assigned (see Fig.~\ref{fig:retrieval_and_clustering} left). In the retrieval setting, the goal is to retrieve \textit{unlabelled} instances based on a set of patient attributes. To that end, we exploit each clinical prototype as a query to search through an \textit{unlabelled} database and retrieve instances to which it is most similar (see Fig.~\ref{fig:retrieval_and_clustering} right).

In designing clinical prototypes, we take inspiration from the field of natural language processing (NLP) where a learnable word embedding represents a unique word. In our case, each prototype represents a unique combination of discrete patient attributes. Formally, given the attributes, $\alpha_{c}$, $\alpha_{s}$, and $\alpha_{a}$, we would have $M = |\alpha_{c}| \times |\alpha_{s}| \times |\alpha_{a}|$ such unique combinations denoted by $A = \{\alpha_{c}^{j},\alpha_{s}^{j},\alpha_{a}^{j}\}_{j=1}^{M}$. We associate each combination, $A_{j} \in A$, with a learnable prototype, $\boldsymbol{p_{A_{j}}} \in \mathbb{R}^{E}$, for a set of $M$ prototypes, $P = \{\boldsymbol{p_{A_{j}}}\}_{j=1}^{M}$. Note that this framework extends to any number of attributes. In the next section, we outline how to learn these clinical prototypes.

\subsection{Learning attribute-specific clinical prototypes}

Clinical prototypes will serve a dual purpose of attribute-specific clustering and retrieval. As such, prototypes will need to be in proximity to a subset of representations of instances associated with a specific set of patient attributes. To achieve this proximity, we leverage the contrastive learning framework which involves a sequence of attractions and repulsions, as explained next.

\paragraph{Hard assignment} We encourage the representation, $\boldsymbol{v_{i}}=f_{\theta}(\boldsymbol{x_{i}})$, of an instance, $\boldsymbol{x_{i}}$, associated with a set of attributes, $A_{i} \in A$, to be similar to the \textit{single} clinical prototype, $\boldsymbol{p_{A_{j}}}$, that shares the exact same set of attributes (i.e., $A_{i} = A_{j}$), and dissimilar to the remaining clinical prototypes, $\{ \boldsymbol{p_{A_{k}}} \}_{k \neq j}$. To achieve this, we optimize $\mathcal{L}_{NCE-hard}$ for a mini-batch of $B$ instances (Eq.~\ref{eq:hard_contrastive}). Intuitively, it heavily penalizes the learner if less probability mass is placed on the similarity of $\boldsymbol{v_{i}}$ and $\boldsymbol{p_{A_{j}}}$ than on the similarity of other representation-and-prototype pairs. We choose to quantify the cosine similarity of pairs, $s(\boldsymbol{v_{i}},\boldsymbol{p_{A_{j}}})$, alongside a temperature parameter, $\tau_{s}$, as is done by Kiyasseh \textit{et al.} \cite{Kiyasseh2020CLOCS}. 
\begin{equation}
    \mathcal{L}_{NCE-hard} = - \frac{1}{B} \sum_{i=1}^{B} \log \left(\frac{e^{s(\boldsymbol{v_{i}},\boldsymbol{p_{A_{j}}})}}{\sum_{k}^{M} e^{s(\boldsymbol{v_{i}},\boldsymbol{p_{A_{k}}})}} \right)
    \qquad \qquad
    s(\boldsymbol{v_{i}},\boldsymbol{p_{A_{j}}}) = \frac{\boldsymbol{v_{i}} . \boldsymbol{p_{A_{j}}}}{\lVert \boldsymbol{v_{i}} \rVert \lVert \boldsymbol{p_{A_{j}}} \rVert} . \frac{1}{\tau_{s}}
    \label{eq:hard_contrastive}
\end{equation}
We refer to this many-to-one mapping from representations to clinical prototype as a hard assignment. Such an assignment, however, implies that a prototype is unlikely to extract potentially useful information from a representation whose attributes are not \textit{exactly} the same as those of the prototype. We quantify this limitation in Sec.~\ref{sec:marginal_impact} and propose an alternative assignment next.

\begin{figure*}[!h]
    \centering
    \begin{subfigure}{1.0\textwidth}
    \includegraphics[width=1\textwidth]{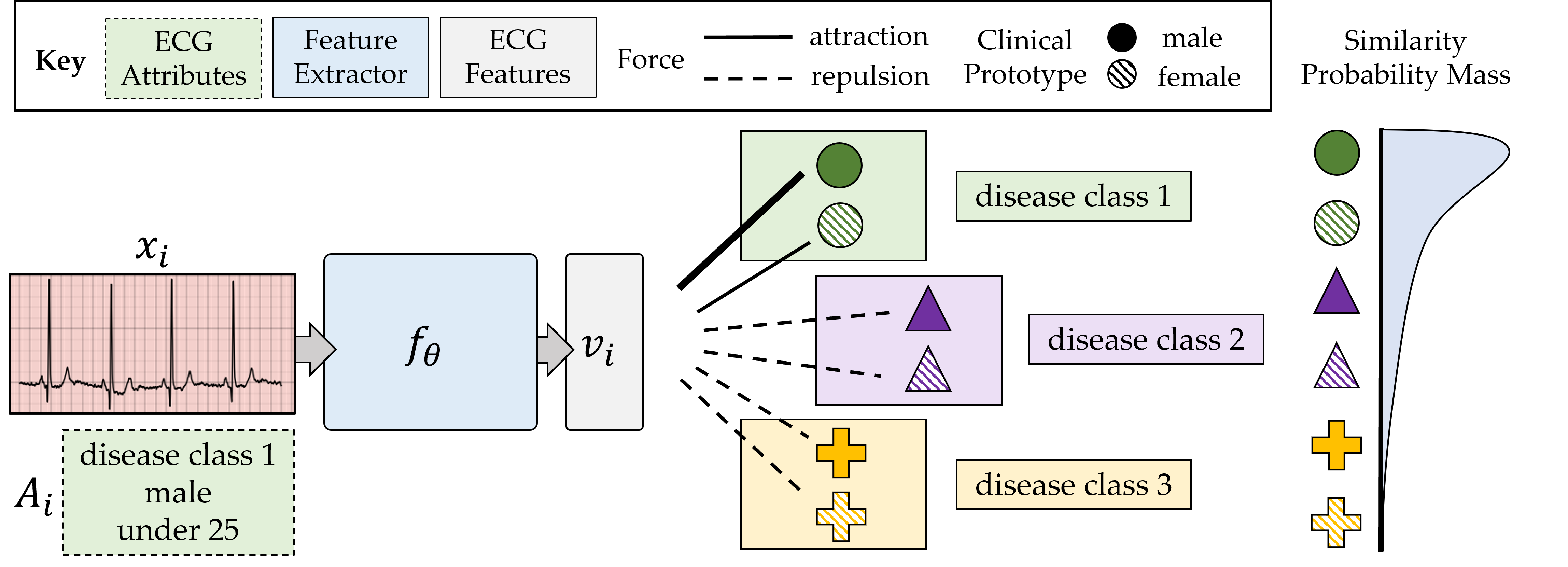}
    \end{subfigure}
    \caption{\textbf{Clinical prototypes are learned via a supervised contrastive learning framework referred to as CROCS.} The representation, $\boldsymbol{v_{i}}$, of an instance, $\boldsymbol{x_{i}}$, associated with a set of attributes, $A_{i}$, is strongly attracted to the clinical prototype which represents the same attributes, weakly attracted to others within the same disease class (colour), and repelled from those representing different classes. These attractions result in the shown similarity probability mass function. To avoid clutter, we have omitted the age attribute associated with clinical prototypes.}
    \label{fig:pipeline}
\end{figure*}

\paragraph{Soft assignment} To overcome the limitation of a hard assignment, we encourage the representation, $\boldsymbol{v_{i}}$, to be similar to a \textit{subset} of clinical prototypes, $L \subset P$ (see Fig.~\ref{fig:pipeline}). We must take caution, however, to avoid erroneously attracting representations to clinical prototypes from a \textit{different} class. Doing so would reduce class-specific margins and thus hinder the downstream clustering and retrieval tasks. 

\textit{Uniform attraction.} We chose the subset to include prototypes, $L = \{ \boldsymbol{p_{A_{l}}} \}_{l=1}^{|L|}$, that share the same disease class, $\alpha_{c}^{i}$, with the representation, $\boldsymbol{v_{i}}$, implying that $A_{l} \in \{\alpha_{c}^{i},\alpha_{s}^{l},\alpha_{a}^{l}\}_{l=1}^{|L|} \subset A$. Note that the clinical prototypes in the subset, $L$, continue to represent attributes that vary along the dimensions of patient sex and age ($\alpha_{s}$, $\alpha_{a}$). Therefore, attracting $\boldsymbol{v_{i}}$ to prototypes in $L$ \textit{uniformly} will likely cause the latter to become minimally distinguishable across sex and age. This is an undesired outcome in light of our goal of learning \textit{attribute-specific} prototypes. We avoid this behaviour by modulating the degree of attraction between $\boldsymbol{v_{i}}$ and all prototypes in the set, $P$, as outlined next.

\begin{figure*}[!h]
    \centering
    \begin{subfigure}{0.9\textwidth}
    \includegraphics[width=1\textwidth]{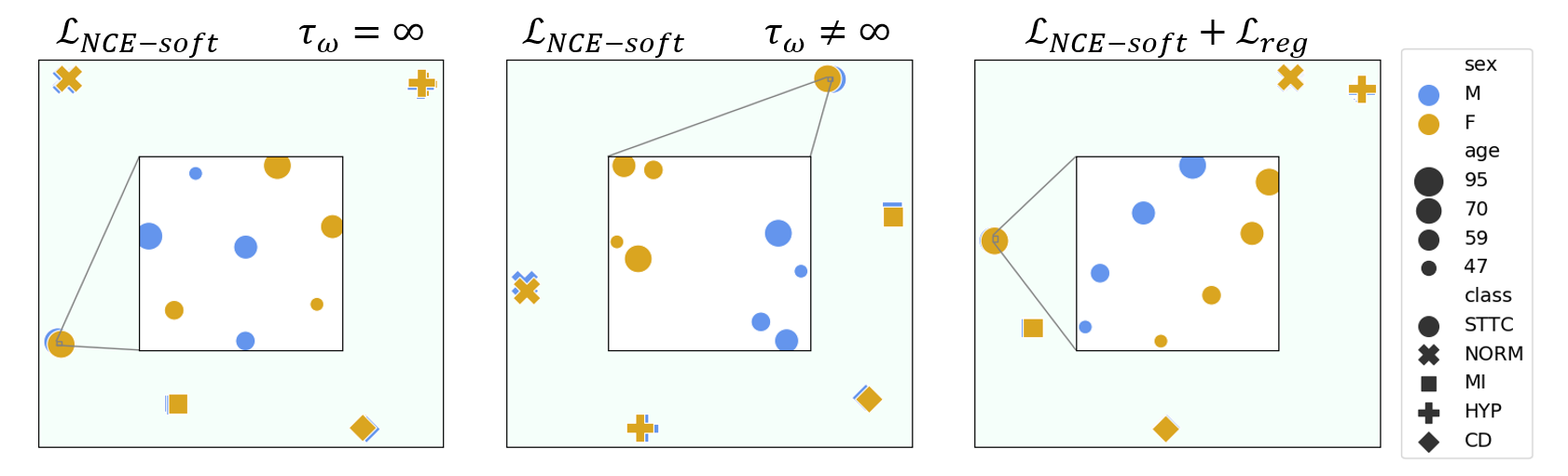}
    \end{subfigure}
    \caption{\textbf{Projection of the set of clinical prototypes, $\boldsymbol{P}$, learned with variants of the CROCS framework.} These prototypes are derived through empirical experiments and projected onto a lower-dimensional space via UMAP \cite{Mcinnes2018}. The legend indicates the attributes of sex (colour), age (size of marker), and disease class (marker symbol). With $\mathcal{L}_{NCE-soft}$ $\tau_{\omega}=\infty$, representations are attracted to class-specific prototypes \textit{uniformly}. With $\mathcal{L}_{NCE-soft}$ $\tau_{\omega}\neq\infty$, prototypes become more linearly separable across sex. Our full framework, $\mathcal{L}_{NCE-soft} + \mathcal{L}_{reg}$, leads to prototypes that adopt a semantically meaningful arrangement. We investigate the marginal impact of these design choices on performance in Sec.~\ref{sec:marginal_impact}.}
    \label{fig:intuition}
\end{figure*}

\textit{Modulated attraction.} The attractive force between $\boldsymbol{v_{i}}$ and $\boldsymbol{p_{A_{j}}}$ is reflected by the corresponding $\mathcal{L}_{NCE-hard}$ term (Eq.~\ref{eq:hard_contrastive}). By placing less probability mass on $s(\boldsymbol{v_{i}},\boldsymbol{p_{A_{j}}})$ (i.e., less similarity) than on $s(\boldsymbol{v_{i}},\boldsymbol{p_{A_{k}}}) \ \forall k, k \neq j$, the learner incurs a higher loss and thus attracts the pair. We extend this logic to all prototypes to obtain $M$ $\mathcal{L}_{NCE-hard}$ terms per representation. To modulate these attractions, we introduce a weight, $w_{ij} \in \{ w_{ij} \}_{j=1}^{M}$, as a coefficient of the $j$-th loss term (Eq.~\ref{eq:soft_contrastive_main}). Each weight, $w_{ij}$, quantifies the degree of matching between attributes of the representation, $A_{i} = \{\alpha_{c}^{i},\alpha_{s}^{i},\alpha_{a}^{i}\}$, and those of the clinical prototype, $A_{j} = \{\alpha_{c}^{j},\alpha_{s}^{j},\alpha_{a}^{j}\}$, as reflected by $q(A_{i},A_{j}) \in \mathbb{R}$. We define $\mathbbm{1}$ as the indicator function and $\tau_{\omega}$ as a temperature parameter that determines how soft the representation-and-prototype attraction is. For example, as $\tau_{\omega} \xrightarrow{} \infty$, this approach reverts to the uniform attraction setup. The intuition is that a stronger attraction ($\uparrow w_{ij}$) should exist for a representation-and-prototype pair that shares more attributes. We also avoid the erroneous attraction of pairs from different classes (i.e., $\alpha_{c}^{i} \neq \alpha_{c}^{j}$) by setting $\omega_{ij}=0$. When visualizing the UMAP projection \cite{Mcinnes2018} of prototypes learned with a uniform attraction ($\tau_{\omega} = \infty$) (Fig.~\ref{fig:intuition} left) and those learned with a modulated attraction ($\tau_{\omega} \neq \infty$) (Fig.~\ref{fig:intuition} centre), we show that the latter become more linearly separable across sex.

\begin{equation}
    \mathcal{L}_{NCE-soft} = - \frac{1}{B} \sum_{i=1}^{B} 
    \left[ \sum_{j=1}^{M} \omega_{ij} \log \left(\frac{e^{s(\boldsymbol{v_{i}},\boldsymbol{p_{A_{j}}})}}{\sum_{k}^{M} e^{s(\boldsymbol{v_{i}},\boldsymbol{p_{A_{k}}})}} \right) \right]
    \label{eq:soft_contrastive_main}
\end{equation}
\begin{equation*}
    \label{eq:weights_main}
    \omega_{ij} = \begin{cases} \frac{e^{q(A_{i},A_{j})}}{\sum_{l}^{|L|} e^{q(A_{i},A_{j})}}& \mbox{if   }     \alpha_{c}^{i}=\alpha_{c}^{j} \\
    0 & \mbox{otherwise}
    \end{cases}
\end{equation*}
\begin{equation*}
    \label{eq:discrepancy_main}
    q(A_{i},A_{j})  = \frac{1}{\tau_{\omega}} . [ \mathbbm{1}(\alpha_{c}^{i}=\alpha_{c}^{j}) + 
     \mathbbm{1}(\alpha_{s}^{i}=\alpha_{s}^{j}) + 
     \mathbbm{1}(\alpha_{a}^{i}=\alpha_{a}^{j}) ]
\end{equation*}

\paragraph{Arrangement of clinical prototypes} Clinical prototypes would confer a high degree of interpretability if they also captured the semantic relationships between attributes. Concretely, prototypes representing similar attribute sets (e.g., adjacent age groups) should be similar to one another. This is analogous to the high similarity of word embeddings representing semantically similar words \cite{Smeaton1999}. To capture these semantic relationships, we encourage class-specific prototypes to maintain some desired distance between one another. As such, each pair of $M$ clinical prototypes, $\boldsymbol{p_{A_{j}}}, \ \boldsymbol{p_{A_{k}}} \ \forall j,k \in [1,M]$ is associated with an empirical and ground-truth (desired) distance. For the former, we normalize the prototypes ($L_{2}$ norm) and calculate their Euclidean distance, $\hat{d}_{jk} = \norm{ \boldsymbol{p_{A_{j}}} - \boldsymbol{p_{A_{k}}} }_{2} \ \forall j,k \in [1,M]$. For the latter, we define the ground-truth distance as $d_{jk} = \beta \times d_{H} \in \mathbb{R}$, where $d_{H}(A_{j},A_{k}) \in \mathbb{Z}^{+}$ is the Hamming distance between a pair of discrete attribute sets. Intuitively, the Hamming distance counts the number of attribute mismatches and $\beta \in \mathbb{R}$ penalizes each mismatch. Therefore, we can generate an \textit{empirical} set, $\{ \hat{d}_{jk} \}_{j,k=1}^{M}$ and a \textit{ground-truth} set, $\{ d_{jk} \}_{j,k=1}^{M}$, of distance values. By minimizing the mean-squared error between these two sets, we learn clinical prototypes that adopt a semantically meaningful arrangement (see Fig.~\ref{fig:intuition} right). Since we are only interested in adopting this arrangement for prototypes of the same class (i.e., $\alpha_{c}^{j} = \alpha_{c}^{k}$), we incorporate the regularization term, $\mathcal{L}_{reg}$, into the final objective function, $\mathcal{L}_{tot}$. 
\begin{equation}
    \mathcal{L}_{reg} = \sum_{j,k=1}^{M} (\hat{d}_{jk} - d_{jk})^{2} \Leftrightarrow \alpha_{c}^{j} = \alpha_{c}^{k} 
    \qquad \qquad
    \mathcal{L}_{tot} = \mathcal{L}_{NCE-soft} + \mathcal{L}_{reg}
    \label{eq:full_loss}
\end{equation}

\section{Experimental design}
\label{sec:design}

\paragraph{Datasets} We use 1) \textbf{Chapman} \cite{Zheng2020} which consists of 12-lead ECG recordings from 10,646 patients alongside cardiac arrhythmia (disease) labels which we group into 4 major classes. 2) \textbf{PTB-XL} \cite{Wagner2020} consists of 12-lead ECG recordings from 18,885 patients alongside disease labels which we group into 5 major classes \cite{Strodthoff2020}. Each dataset contains patient sex and age information and is split, at the patient level, into training, validation, and test sets using a $60:20:20$ configuration. Each time-series recording is split into non-overlapping segments of $2500$ samples ($\approx 5$s in duration), as this is common for in-hospital recordings. Further details are provided in Appendix~\ref{appendix:datasets}.

\paragraph{Description of clustering setting} During inference, we treat the clinical prototypes, $\{\boldsymbol{p_{A_{j}}}\}_{j=1}^{M}$, as a set of cluster centroids. We calculate the Euclidean distance between the $i$-th representation and each of the $M$ prototypes, identify the closest prototype, $\boldsymbol{p_{A_{k}}}$, and assign the representation to $A_{k}$ which we now denote by $\hat{A}_{i} = \{\hat{\alpha}_{c}^{i},\hat{\alpha}_{s}^{i},\hat{\alpha}_{a}^{i} \}$. Repeating this process for $N$ unseen instances results in a set of assigned attribute values, $\vec{\hat{\alpha}} = \{ \hat{\alpha}^{i} \}_{i=1}^{N}$, for a particular attribute, $\hat{\alpha} \in A$ (e.g., disease class). Such unseen instances would typically be \textit{unlabelled}. For evaluation, however, we assume access to the ground-truth attribute values, $\vec{\alpha}=\{ \alpha^{i} \}_{i=1}^{N}$, with which we calculate the accuracy, $\text{Acc}(\hat{\alpha})$, and the adjusted mutual information, $\text{AMI}(\hat{\alpha}) \in [0,1]$, between $\vec{\hat{\alpha}}$ and $\vec{\alpha}$. 
\begin{equation}
    \text{Acc}(\hat{\alpha}) = \frac{1}{N} \sum_{i=1}^{N} \mathbbm{1}(\hat{\alpha}^{i} = \alpha^{i})
    \qquad \qquad 
    \text{AMI}(\hat{\alpha}) = \frac{\left[\mathbbm{MI}(\vec{\alpha},\vec{\hat{\alpha}}) - \mathbb{E}(\mathbbm{MI}(\vec{\alpha},\vec{\hat{\alpha}})) \right]}{\mathbb{E}(\mathbbm{H}(\vec{\alpha}),\mathbbm{H}(\vec{\hat{\alpha}})) - \mathbb{E}(\mathbbm{MI}(\vec{\alpha},\vec{\hat{\alpha}}))}
    \label{eq:ami}
\end{equation}
where $\mathbbm{MI}(\vec{\alpha},\vec{\hat{\alpha}})$ denotes the mutual information between the ground-truth and assigned set of attribute values, and $\mathbbm{H}(\vec{\alpha})$ denotes the entropy of the attribute values.

\paragraph{Description of retrieval setting} During inference, we treat the clinical prototypes, $\{\boldsymbol{p_{A_{j}}}\}_{j=1}^{M}$, as a query set. We calculate the Euclidean distance between the $j$-th clinical prototype and representations of $N$ unseen instances, retrieve the $K$ closest instances, and then assign them to $A_{j} = \{ \alpha_{c}^{j}, \alpha_{s}^{j}, \alpha_{a}^{j} \}$. Note that such instances would typically be \textit{unlabelled}, thus precluding a simple SQL search. For evaluation, however, we assume access to the ground-truth attributes, $\{ \alpha_{c}^{i}, \alpha_{s}^{i}, \alpha_{a}^{i} \}_{i=1}^{K}$, with which we calculate a variant of the precision at $K$ metric (Eq.~\ref{eq:recall_at_k}). It checks whether at least one of the retrieved instances is relevant, where relevance is based on a partial or exact match of query and instance attributes (\# attribute matches). This value is then averaged across all $M$ prototypes. 
\begin{equation}
    \text{P\scriptsize{@}}K = \frac{1}{M} \sum_{j=1}^{M} \mathbbm{1} \left ( \sum_{i=1}^{K} \mathbbm{1} \underbrace{ \left ( \left [ \alpha_{c}^{i} = \alpha_{c}^{j} \right ] \cap \left [ \alpha_{s}^{i} = \alpha_{s}^{j} \right ] \cap \left [ \alpha_{a}^{i} = \alpha_{a}^{j} \right ] \right )}_{\text{relevance $\equiv$ \# attribute matches = 3} }  \geq 1 \right )
    \label{eq:recall_at_k}
\end{equation}
Ideally, a retrieved instance would match all of the query's attributes (disease class, sex, and age). In our context, however, the motivation for breaking down the evaluation of the retrieval setting based on the number of attribute matches is twofold. First, it allows us to evaluate our framework at a more granular level and thus detect subtle changes in performance. Second, the importance of such an evaluation at the attribute level arguably depends on the clinical context. For example, a cardiologist diagnosing a disease might be most interested in the pathology (i.e., disease class) attribute. On the other hand, a pharmaceutical company looking to recruit patients within a particular age group for a clinical trial might be most interested in the age attribute. Moreover, biomedical researchers looking to stratify treatment outcomes based on sex would be most interested in the sex attribute.

\paragraph{Baseline methods} We compare clinical prototypes learned via the CROCS framework (\textbf{CP CROCS}) to the following methods. For retrieval, \textbf{Deep Transfer Cluster (DTC)} \cite{Han2019} learns cluster prototypes by minimizing the KL divergence between a target distribution and one based on the distance between prototypes and representations. \textbf{TP CROCS} involves traditional prototypes where each prototype, $\boldsymbol{\bar{v}_{A_{j}}} = \frac{1}{\sum \mathbb{I}(A_{i}=A_{j}}) \sum_{i=1}^{N} \boldsymbol{v_{i}} \cdot \mathbb{I}(A_{i}=A_{j})$, is simply an average of representations, $\boldsymbol{v_{i}}$, associated with the same set of attributes, $A_{j}$. Such representations are also learned via CROCS.

For the clustering task, we compare to several state-of-the-art clustering methods, in addition to those mentioned above. $k$-means identifies cluster centroids based on the input instances, $\boldsymbol{x}$ (\textbf{KM raw}), or representations, $\boldsymbol{v}$, learned via the CROCS (\textbf{KM CROCS}) or explainable prototypes (\textbf{KM EP}) \cite{Gee2019} framework. \textbf{DeepCluster (DC)} \cite{Caron2018} iteratively applies $k$-means to representations, pseudo-labels them according to their assigned cluster, and then exploits such labels for supervised training. \textbf{Deep Temporal Clustering Representation (DTCR)} \cite{Ma2019} optimizes an objective function with a reconstruction, $k$-means, and classifier loss that determines whether instances are real. \textbf{Information Invariant Clustering (IIC)} \cite{Ji2019} maximizes the mutual information between class probabilities of an instance and its perturbed counterpart. \textbf{SeLA} \cite{Asano2020} implements Sinkhorn-Knopp to pseudo-label instances in a supervised manner. For further details, see Appendix~\ref{appendix:baseline_methods}.


\paragraph{Hyperparameters} During optimization, we chose the temperature parameter, $\tau_{s}=0.1$ \cite{Kiyasseh2020CLOCS}, and $\tau_{w}=1$. We specified $\beta=0.2$, in the regularization term, after experimenting with several values (see Appendix~\ref{appendix:effect_of_beta}). Too small a value of $\beta$ would decrease the distance between class-specific clinical prototypes. Too large a value of $\beta$ would cause clinical prototypes from \textit{different} classes to overlap with one another and thus reduce class separability. For Chapman and PTB-XL, $\mathrm{sex} \in \{\mathrm{M},\mathrm{F}\}$, $\mathrm{age}$ is converted to quartiles, and $|\mathrm{class}|=4$ and $5$, respectively. Therefore, $M=|\mathrm{class}| \times |\mathrm{sex}| \times |\mathrm{age}|=32$ and $40$, for the two datasets, respectively. The network, $f_{\theta}$, comprises 1D convolutional operators and we chose the embedding dimension $E=128$ and $256$ for Chapman and PTB-XL, respectively. We show that $E$ has a minimal effect on performance (see Appendix~\ref{appendix:effect_of_embedding_size}). Further network and implementation details can be found in Appendix~\ref{appendix:network}. 

\section{Experimental results}

\subsection{Visualizing clinical prototypes}

\begin{figure}[!h]
    \centering
    \begin{subfigure}{0.48\textwidth}
    \centering
    \includegraphics[width=1\textwidth]{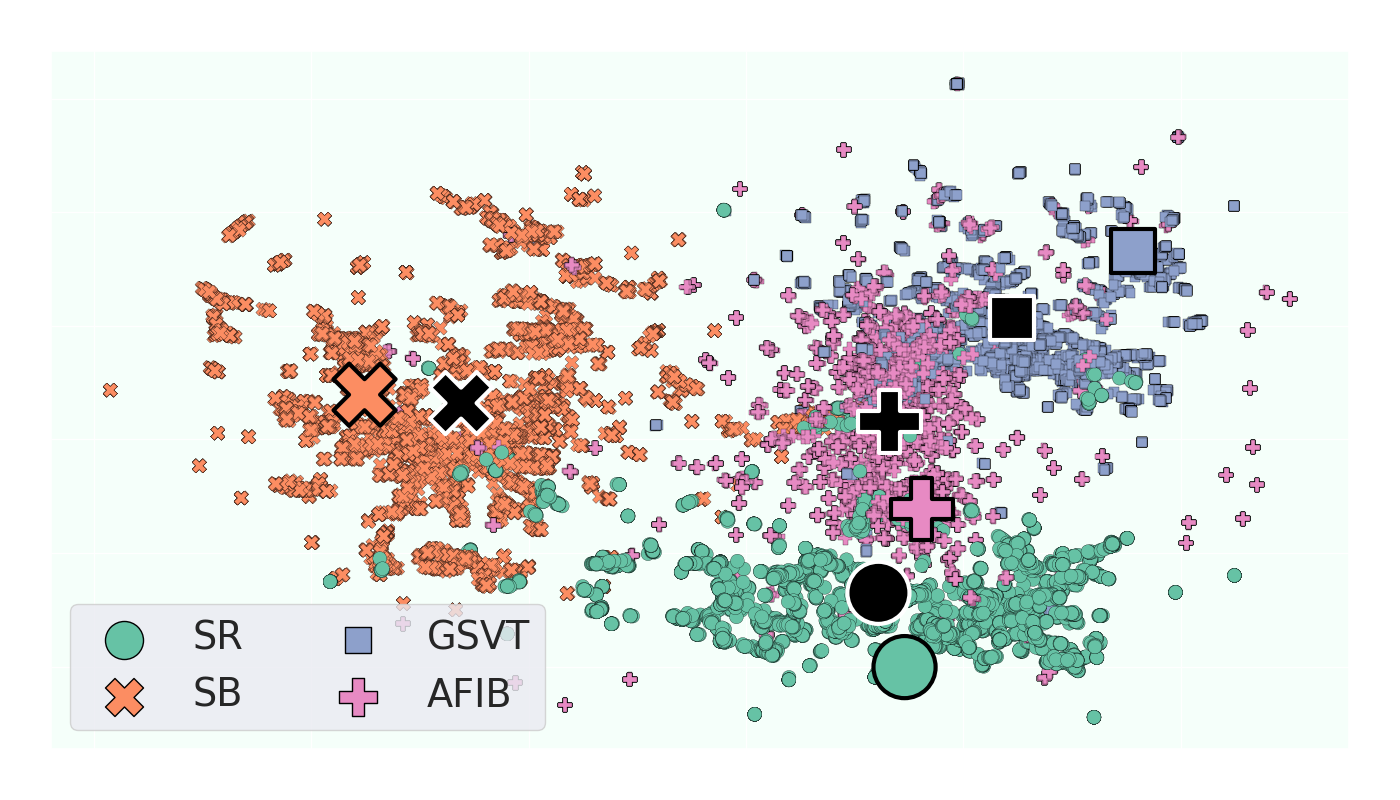}
    \end{subfigure}
    ~
    \begin{subfigure}{0.48\textwidth}
    \centering
    \includegraphics[width=1\textwidth]{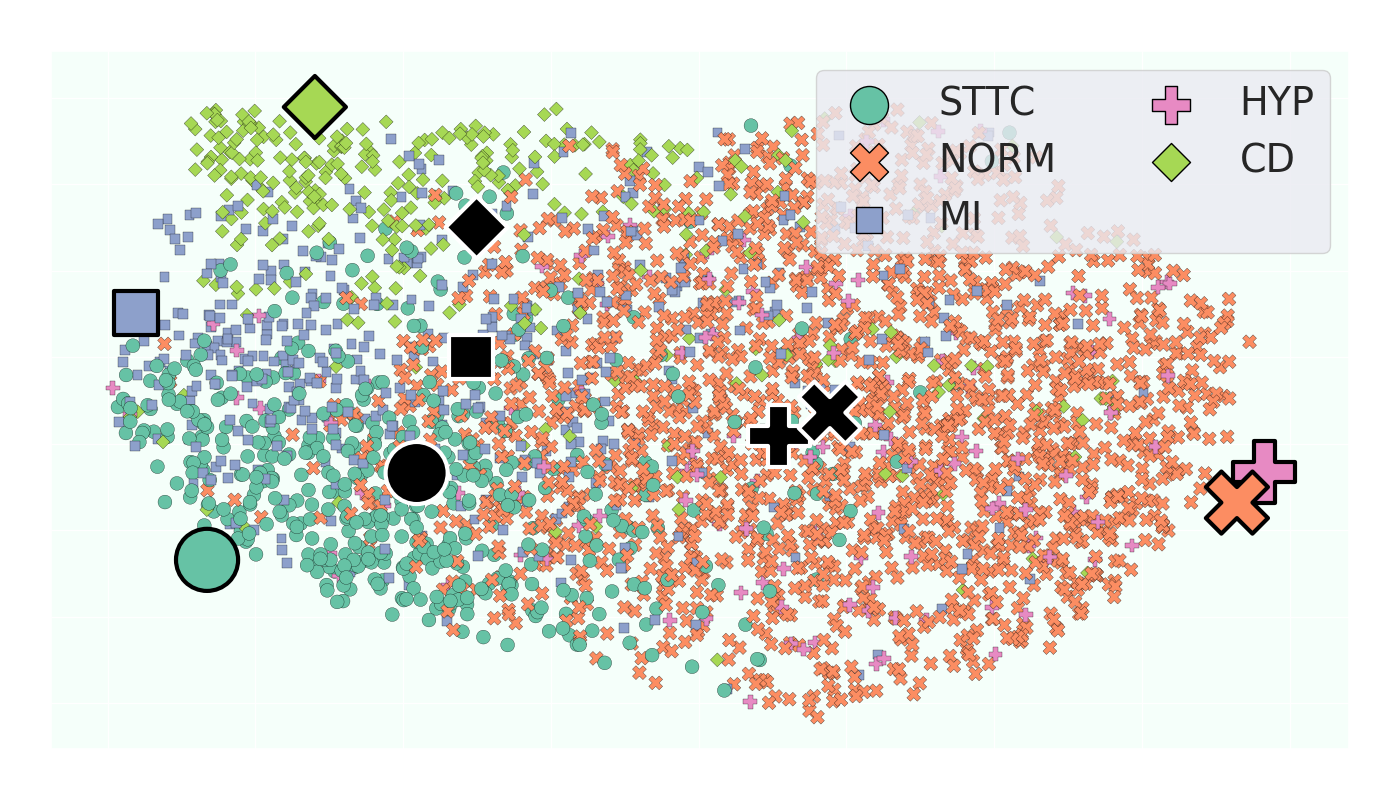}
    \end{subfigure}
    \caption{\textbf{Projection of class-specific clinical prototypes $\boldsymbol{p}$ (large, coloured shapes), traditional prototypes $\boldsymbol{\bar{v}}$, (large, black shapes), and representations, $\boldsymbol{v}$, in the validation set of (left) Chapman and (right) PTB-XL.} We show that clinical prototypes are class-specific, consistent with the class labels of representations, and distinct from traditional prototypes. This bodes well for their use as centroids for clustering (see Sec.~\ref{sec:clustering}) and as queries for retrieval (see Sec.~\ref{sec:retrieval}).}
    \label{fig:tsne_figure}
\end{figure}

We begin by qualitatively validating the claim that clinical prototypes are attribute-specific. In other words, can prototypes be delineated along the dimensions of disease class, sex, and age? To address this, we illustrate, in Fig.~\ref{fig:tsne_figure}, two dimensional UMAP projections of the class-specific clinical prototypes (large, coloured shapes), traditional prototypes (large, black shapes), and representations of instances in the validation set of Chapman and PTB-XL. 

We show that clinical prototypes are indeed disease class-specific, as evident by the high degree of class separability of such prototypes. We also find that clinical prototypes are distinct from traditional prototypes, a distinction whose importance will become evident in later sections. Second, the consistency of the class labels of the prototypes with those of the representations is a harbinger of how prototypes may perform in the clustering and retrieval settings, as we show in the next section. These findings complement the delineation of the prototypes along the dimensions of sex and age, and their adoption of a semantically meaningful arrangement, as was shown in Section~\ref{sec:methods}.


\subsection{Deploying clinical prototypes in clustering setting}
\label{sec:clustering}

In the clustering setting, we assign cardiac signals in a held-out dataset to a set of patient attributes associated with the cluster of the closest clinical prototype. We evaluate these assignments based on the three patient-specific attributes (disease class, sex, and age) and present the results in Table~\ref{table:clustering_results}.

\begin{table}[!h]
    \centering
    \small
    \caption{\textbf{Clustering performance on the validation set of Chapman and PTB-XL.} Evaluation is based on (a) class and (b) sex and age attributes. Results are averaged across five random seeds. Brackets indicate standard deviation and bold reflects the top-performing method. We show that $\mathrm{CP \ CROCS}$ outperforms the remaining methods regardless of patient attribute.}
    \label{table:clustering_results}
    \begin{subtable}{0.48\textwidth}
    \caption{Cardiac arrhythmia class attribute}
    \label{table:clustering_results_class}
    \centering
    \resizebox{\textwidth}{!}{
    \begin{tabular}{c | c c | c c}
        \toprule
         \multirow{2}{*}{Method} & \multicolumn{2}{c}{Chapman} & \multicolumn{2}{c}{PTB-XL} \\
         & Acc & AMI & Acc & AMI \\ 
        \midrule

         SeLA \cite{Asano2020} & $21.0$ {\scriptsize(0.1)} & $9.2$ {\scriptsize(10.0)} & $10.5$ {\scriptsize(0.1)} & $1.6$ {\scriptsize(0.5)} \\ 
         
         DC \cite{Caron2018} & $21.0$ {\scriptsize(0.1)} & $3.0$ {\scriptsize(4.0)} & $10.5$ {\scriptsize(0.1)} & $4.9$ {\scriptsize(0.0)} \\
         
         IIC \cite{Ji2019} & $27.0$ {\scriptsize(0.2)} & $0.2$ {\scriptsize(0.0)} & $22.0$ {\scriptsize(0.7)} & $0.5$ {\scriptsize(0.0)} \\
        
         DTCR \cite{Ma2019} & $29.3$ {\scriptsize(1.3)} & $0.2$ {\scriptsize(0.2)} & 38.4 {\scriptsize(4.2)} & $1.4$ {\scriptsize(0.3)} \\
        
         DTC \cite{Han2019} & $53.4$ {\scriptsize(15.0)} & $23.4$ {\scriptsize(20.0)} & $67.3$ {\scriptsize(1.0)} & $25.2$ {\scriptsize(0.7)} \\
         
         KM raw & $28.4$ {\scriptsize(1.2)} & $0.3$ {\scriptsize(0.0)} & - & - \\
         
         KM EP \cite{Gee2019} & $65.6$ {\scriptsize(4.0)} & $42.8$ {\scriptsize(2.6)} & $48.9$ {\scriptsize(1.1)} & $24.2$ {\scriptsize(1.9)} \\
         
         KM CROCS & $73.4$ {\scriptsize(7.1)} & $58.6$ {\scriptsize(2.8)} & $47.6$ {\scriptsize(3.9)} & $25.9$ {\scriptsize(1.1)} \\
         
         TP CROCS & $80.3$ {\scriptsize(1.4)} & $65.0$ {\scriptsize(0.6)} & $53.6$ {\scriptsize(0.7)} & $29.1$ {\scriptsize(0.2)} \\

         CP CROCS & $\boldsymbol{90.3}$ {\scriptsize(0.8)} & $\boldsymbol{72.8}$ {\scriptsize(1.6)} & $\boldsymbol{76.0}$ {\scriptsize(0.3)} & $\boldsymbol{35.9}$ {\scriptsize(0.4)} \\
         \bottomrule
    \end{tabular}
    }
    \end{subtable}
    ~
    \begin{subtable}{0.48\textwidth}
    \centering
    \caption{Sex and age attributes}
    \label{table:clustering_results_nonclass}
    \resizebox{\textwidth}{!}{
    \begin{tabular}{c | c c | c c}
        \toprule
         \multirow{2}{*}{Method} & \multicolumn{2}{c}{Chapman} & \multicolumn{2}{c}{PTB-XL} \\
         & sex & age & sex & age \\ 
         \midrule
         
         DTCR \cite{Ma2019} & $51.2$ {\scriptsize(0.9)} & $25.9$ {\scriptsize(0.1)} & $51.0$ {\scriptsize(0.9)} & $25.1$ {\scriptsize(0.8)} \\ 
         
         DTC \cite{Han2019} & $54.8$ {\scriptsize(0.5)} & $26.4$ {\scriptsize(0.8)} & $58.6$ {\scriptsize(1.9)} & $25.7$ {\scriptsize(0.7)} \\

         KM EP \cite{Gee2019} & $56.1$ {\scriptsize(0.0)} & $31.0$ {\scriptsize(0.4)} & $54.1$ {\scriptsize(0.3)} & $29.2$ {\scriptsize(2.0)} \\

         KM CROCS & $54.9$ {\scriptsize(0.8)} & $32.3$ {\scriptsize(0.5)} & $51.8$ {\scriptsize(1.2)} & $31.6$ {\scriptsize(1.5)} \\
         
         TP CROCS & $54.8$ {\scriptsize(1.0)} & $31.1$ {\scriptsize(1.7)} & $69.7$ {\scriptsize(0.8)} & $\boldsymbol{39.4}$ {\scriptsize(0.4)} \\

         CP CROCS & $\boldsymbol{57.4}$ {\scriptsize(1.2)} & $\boldsymbol{38.0}$ {\scriptsize(0.8)} & $\boldsymbol{73.5}$ {\scriptsize(0.6)} & $19.5$ {\scriptsize(0.2)} \\
         \bottomrule
    \end{tabular}
    }
    \end{subtable}
\end{table}

In Table~\ref{table:clustering_results}, we find that CROCS outperforms both generic and domain-specific state-of-the-art clustering methods. For example, on Chapman, $\mathrm{CP \ CROCS}$, $\mathrm{KM \ EP}$, and $\mathrm{DTC}$ achieve $\mathrm{Acc(class)}=90.3$, $65.6$, and $53.4\%$ respectively. Along the dimension of sex, and on PTB-XL, $\mathrm{CP \ CROCS}$ and $\mathrm{DTC}$ achieve $\mathrm{Acc(sex)}=73.5$ and $58.6\%$, respectively. Second, we find that CROCS leads to rich representation learning that facilitates clustering. This is evident when comparing the performance of $k$-means applied to representations that are learned via different methods. For example, on Chapman, $\mathrm{KM \ raw}$, $\mathrm{KM \ EP}$, and $\mathrm{KM \ CROCS}$ achieve $\mathrm{Acc(class)}=28.4$, $65.6$, and $73.4\%$, respectively. We also find that clinical prototypes, when exploited as centroids, are preferable to traditional prototypes, and centroids learned via $k$-means. For example, on PTB-XL, $\mathrm{KM \ CROCS}$, $\mathrm{TP \ CROCS}$, and $\mathrm{CP \ CROCS}$ achieve $\mathrm{Acc(class)}=47.6$, $53.6$, and $76.0\%$, respectively. These findings, which hold across datasets and evaluation metrics, point to the overall utility of the CROCS framework and clinical prototypes for attribute-specific clustering. 

\subsection{Deploying clinical prototypes in the retrieval setting}
\label{sec:retrieval}

Up until now, we have shown that CROCS leads to accurate clustering. In this section, we show that CROCS can also be independently exploited for retrieval. Specifically, a query retrieves the closest $K=[1,5,10]$ previously unseen cardiac signals, and assigns them to its associated set of patient attributes. In Table~\ref{table:retrieval_results}, we evaluate these assignments based on both partial and exact matches of the attributes (\# attribute matches) represented by the query and retrieved cardiac signals.

In Table~\ref{table:retrieval_results}, we find that CROCS outperforms the baseline retrieval method, $\mathrm{DTC}$. For example, on Chapman, at $K=1$, and when \# attribute matches $\geq1$, $\mathrm{CP \ CROCS}$ and $\mathrm{DTC}$ achieve a precision of $95.6$ and $71.9\%$, respectively. This indicates that, on average, $95.6\%$ of the cardiac signals retrieved by the clinical prototypes are relevant. Relevance, in this case, implies that the retrieved cardiac signals share at least one attribute with the query. Such a finding points to the utility of clinical prototypes as queries in the retrieval setting. We also find that CROCS leads to rich representation learning that facilitates retrieval. This is evident by the strong performance of $\mathrm{TP \ CROCS}$ which depends directly on representations learned via our CROCS framework. For example, on PTB-XL, at $K=1$, and when \# attribute matches $\geq1$, $\mathrm{DTC}$, $\mathrm{TP \ CROCS}$, and $\mathrm{CP \ CROCS}$ achieve a precision of $70.0$, $99.0$, and $92.5\%$, respectively. In this particular case, the lower performance of $\mathrm{CP \ CROCS}$ relative to $\mathrm{TP \ CROCS}$ is hypothesized to stem from clinical prototypes acting instead as \textit{archetypes} (extreme representative data points) \cite{Morup2012} which may occasionally hinder retrieval along multiple attributes. Evidence of such extreme embeddings can be found in Fig.~\ref{fig:tsne_figure}. We also show that $\mathrm{CROCS}$ continues to perform well even when provided with only $10\%$ of the labelled training data (see Appendix~\ref{appendix:effect_of_data}).

\begin{table}[!h]
\small
\centering
\caption{\textbf{Precision of $\boldsymbol{K}$ retrieved representations, $\boldsymbol{v}$, in the validation set of Chapman and PTB-XL, that are closest to the query.} Results are shown for partial and exact matches of the attributes (\# attribute matches) represented by the query and retrieved cardiac signals, and are averaged across five random seeds. Brackets indicate standard deviation and bold reflects the top-performing method. The strong performance of $\mathrm{CP \ CROCS}$ provides evidence in support of our CROCS framework.}
    \label{table:retrieval_results}
    \centering
    \vskip 0.05in
    \resizebox{0.8\textwidth}{!}{
    \begin{tabular}{c c | c c c | c c c}
        \toprule
        \# attribute & \multirow{2}{*}{Query} & \multicolumn{3}{c}{Chapman} & \multicolumn{3}{c}{PTB-XL} \\
        matches & & $K=1$ & $5$ & $10$ & $1$ & $5$ & $10$ \\
        \midrule
        
        \multirow{3}{*}{$\geq1$} & DTC \cite{Han2019} & $71.9$ {\scriptsize(0.0)} & $100.0$ {\scriptsize(0.0)} & $100.0$ {\scriptsize(0.0)} & $70.0$ {\scriptsize(0.0)} & $90.0$ {\scriptsize(8.4)} & $100.0$ {\scriptsize(0.0)} \\
        
        & TP CROCS & $91.9$ {\scriptsize(3.2)} & $97.5$ {\scriptsize(2.3)} & $100.0$ {\scriptsize(0.0)} & $\boldsymbol{99.0}$ {\scriptsize(2.0)} & $100.0$ {\scriptsize(0.0)} & $100.0$ {\scriptsize(0.0)} \\
        
        & CP CROCS & $\boldsymbol{95.6}$ {\scriptsize(6.1)} & $100.0$ {\scriptsize(0.0)} & $100.0$ {\scriptsize(0.0)} & $92.5$ {\scriptsize(0.0)} & $100.0$ {\scriptsize(0.0)} & $100.0$ {\scriptsize(0.0)} \\
        
        \midrule 
        \multirow{3}{*}{$\geq2$} & DTC \cite{Han2019} & $25.0$ {\scriptsize(0.0)} & $71.9$ {\scriptsize(9.7)} & $90.0$ {\scriptsize(7.0)} & $22.5$ {\scriptsize(0.0)} & $52.5$ {\scriptsize(16.0)} & $80.5$ {\scriptsize(4.0)}\\
        
        & TP CROCS & $55.0$ {\scriptsize(3.8)} & $79.4$ {\scriptsize(7.6)} & $90.0$ {\scriptsize(0.1)} & $\boldsymbol{71.5}$ {\scriptsize(2.0)} & $94.5$ {\scriptsize(1.0)} & $\boldsymbol{100.0}$ {\scriptsize(0.0)} \\
        
        & CP CROCS & $\boldsymbol{61.3}$ {\scriptsize(10.0)} & $\boldsymbol{86.3}$ {\scriptsize(10.9)} & $\boldsymbol{93.8}$ {\scriptsize(7.9)} & $63.0$ {\scriptsize(1.9)} & $\boldsymbol{96.5}$ {\scriptsize(2.0)} & $99.5$ {\scriptsize(1.0)} \\
        \midrule
        \multirow{3}{*}{$=3$} & DTC \cite{Han2019} & $3.1$ {\scriptsize(0.0)} & $15.0$ {\scriptsize(3.1)} & $23.8$ {\scriptsize(0.1)} & $2.5$ {\scriptsize(0.0)} & $9.5$ {\scriptsize(4.3)} & $16.5$ {\scriptsize(0.2)}\\

        & TP CROCS & $10.6$ {\scriptsize(1.5)} & $23.8$ {\scriptsize(6.1)} & $36.9$ {\scriptsize(7.0)} & $\boldsymbol{15.5}$ {\scriptsize(1.0)} & $32.0$ {\scriptsize(1.0)} & $43.5$ {\scriptsize(0.1)} \\

        & CP CROCS & $\boldsymbol{11.3}$ {\scriptsize(2.5)} & $\boldsymbol{33.1}$ {\scriptsize(6.1)} & $\boldsymbol{46.3}$ {\scriptsize(6.7)} & $12.5$ {\scriptsize(0.0)} & $\boldsymbol{33.5}$ {\scriptsize(2.0)} & $\boldsymbol{43.0}$ {\scriptsize(4.0)} \\
        \bottomrule
\end{tabular}
}
\end{table}

\begin{figure}[!h]
    \centering
    \begin{subfigure}{0.48\textwidth}
    \centering
	\includegraphics[width=1\textwidth]{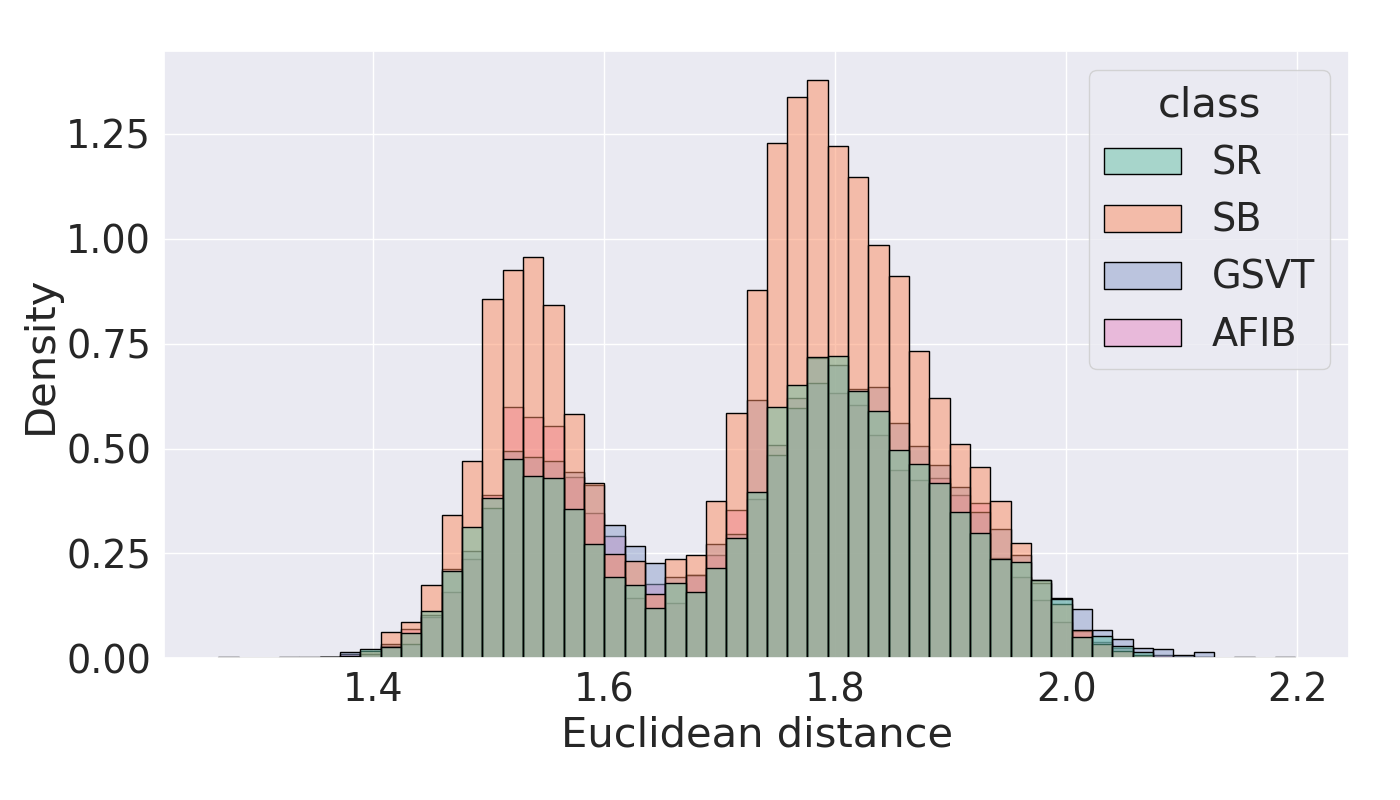}
	\end{subfigure}
	~
    \begin{subfigure}{0.48\textwidth}
    \centering
	\includegraphics[width=1\textwidth]{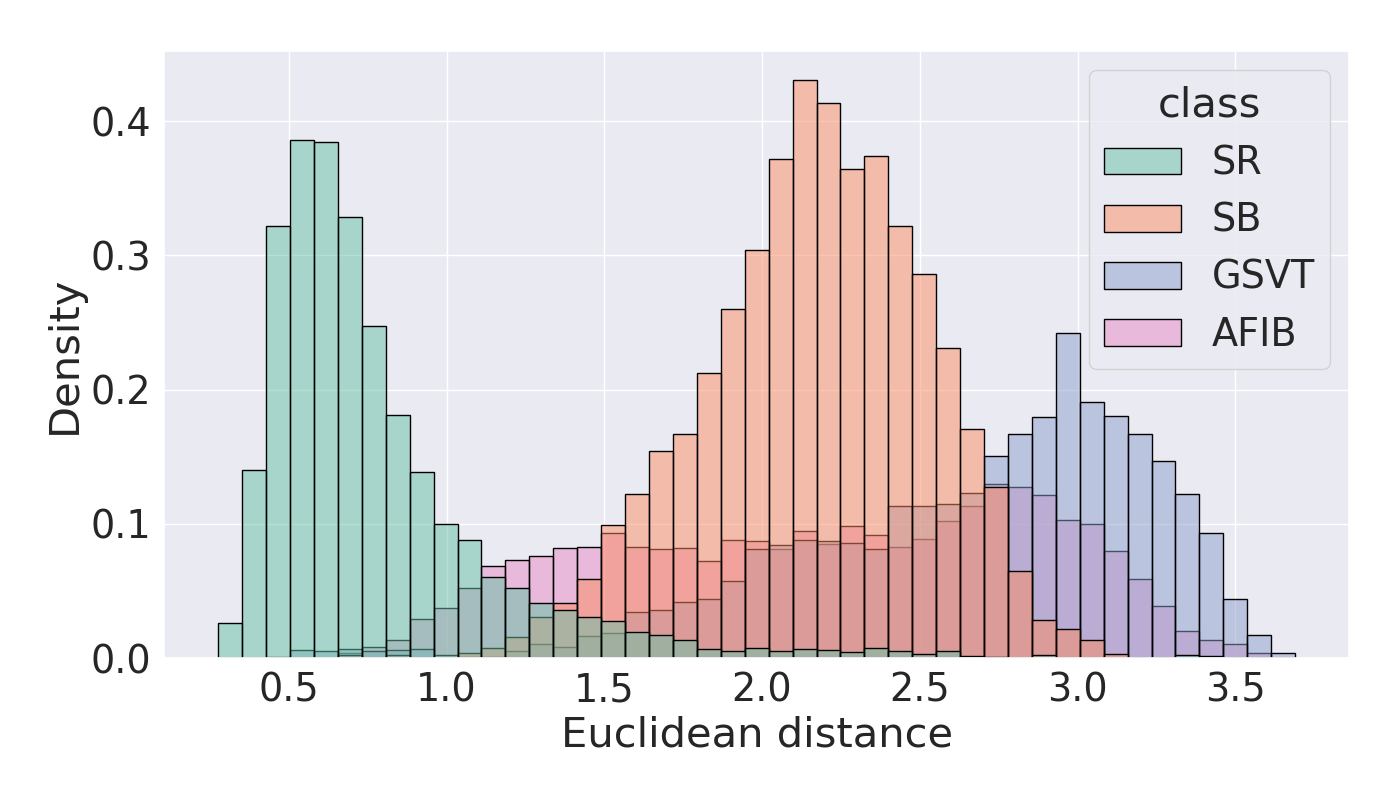}
	\end{subfigure}
	~
    \begin{subfigure}{0.48\textwidth}
    \centering
	\includegraphics[width=1\textwidth]{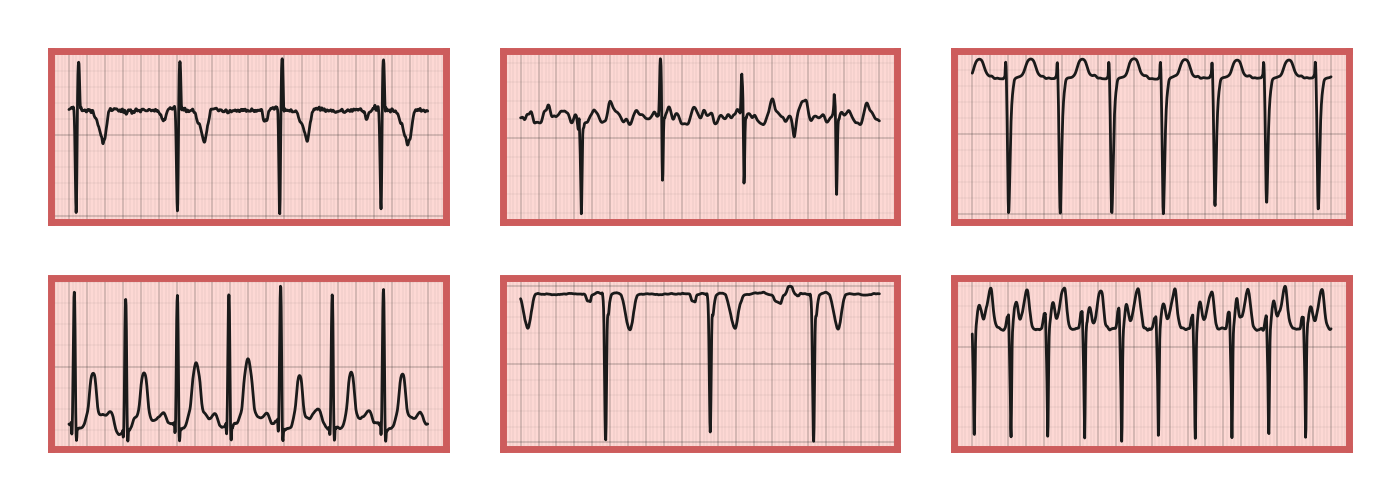}
	\caption{$\mathrm{DTC}$ query $\{\mathrm{SR},\mathrm{male},\mathrm{under \ 49}\}$}
	\label{fig:chapman_distances_dtc}
	\end{subfigure}
	~
    \begin{subfigure}{0.48\textwidth}
    \centering
	\includegraphics[width=1\textwidth]{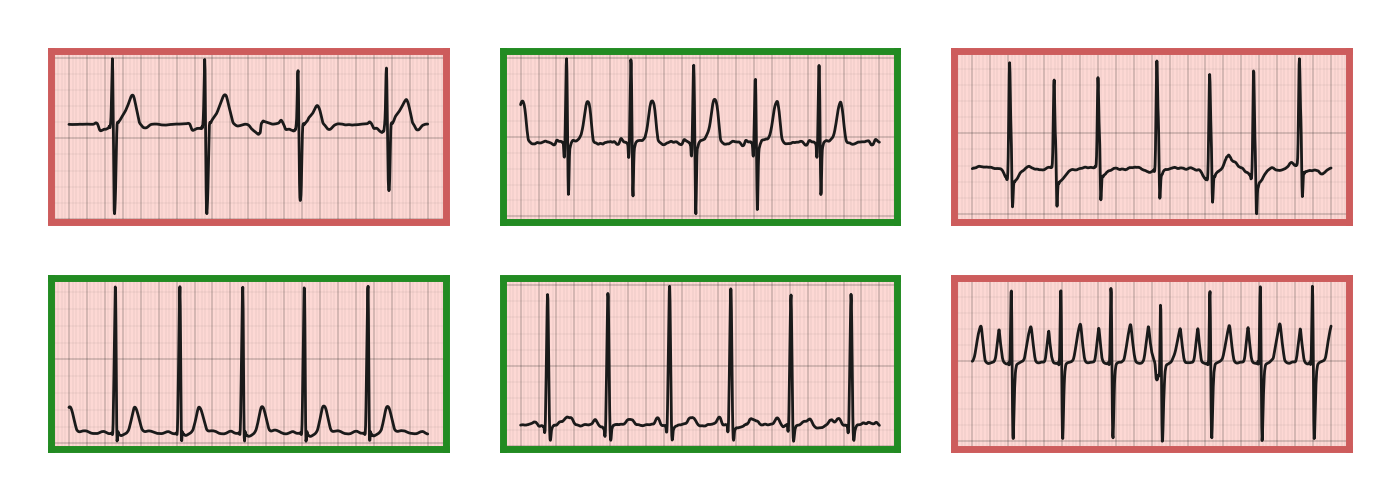}
	\caption{$\mathrm{CP \ CROCS}$ query $\{\mathrm{SR},\mathrm{male},\mathrm{under \ 49}\}$}
	\label{fig:chapman_distances_cp}
	\end{subfigure}
	\caption{\textbf{Qualitative retrieval performance of two distinct query prototypes.} (\textbf{top row}) Euclidean distance from (a) $\mathrm{DTC}$ query or (b) $\mathrm{CP \ CROCS}$ query to representations, $\boldsymbol{v}$, in the validation set of Chapman. (\textbf{bottom row}) Six closest cardiac signals to the query which is associated with a set of patient attributes $\{ \mathrm{disease}, \mathrm{sex}, \mathrm{age} \}$. Retrieved cardiac signals with a green border indicate those whose class attribute matches that of the query. We show that the $\mathrm{CP \ CROCS}$ query is closer to representations of the same class ($\mathrm{SR}$) and thus retrieves relevant cardiac signals.}
	\label{fig:chapman_top_k_ecgs}
\end{figure}

To qualitatively evaluate the retrieval performance, we first randomly choose a query representing a set of attributes and calculate its Euclidean distance to the representations in a validation set. We present distributions of such distance values in Fig.~\ref{fig:chapman_top_k_ecgs} (top row), for a $\mathrm{DTC}$ query and a $\mathrm{CP \ CROCS}$ query, coloured based on the ground-truth class of the representations (other queries shown in Appendix~\ref{appendix:retrieval}). In Fig.~\ref{fig:chapman_top_k_ecgs} (bottom row), we illustrate the six cardiac signals ($K=6$) that are closest to each query, with a green border indicating signals whose class attribute matches that of the query. We find that the $\mathrm{CP \ CROCS}$ query is closer to representations of the same class ($\mathrm{SR}$) than to those of a different class. For example, in Fig.~\ref{fig:chapman_distances_cp} (top row), the average distance between the $\mathrm{CP \ CROCS}$ query representing $\{\mathrm{SR},\mathrm{male},\mathrm{under \ 49}\}$ and representations with and without the class attribute $\mathrm{SR}$ is $\approx 0.6$ and $>1.5$, respectively. Such separability, which is not exhibited by the $\mathrm{DTC}$ query, points to the improved reliability of the $\mathrm{CP \ CROCS}$ query in distinguishing between the relevance of cardiac signals. Further evidence in support of this reliability is shown in Fig.~\ref{fig:chapman_top_k_ecgs} (bottom row) where we find that a $\mathrm{DTC}$ and a $\mathrm{CP \ CROCS}$ query retrieve relevant cardiac signals $0\%$ and $50\%$ of the time, respectively. This finding also extends to the PTB-XL dataset (Appendix~\ref{appendix:retrieval}). 

\begin{table}[!b]
    \centering
    \small
    \caption{\textbf{Marginal impact of design choices of CROCS on clustering performance.} Evaluation is based on (a) class and (b) sex and age attributes. Results are averaged across five random seeds. Brackets indicate standard deviation and bold reflects the top-performing method. We show that our full framework ($\mathcal{L}_{NCE-soft} + \mathcal{L}_{reg}$) is preferable to other variants regardless of attribute.}
    \label{table:ablation_clustering_results}
    \begin{subtable}{0.45\textwidth}
    \caption{Cardiac arrhythmia class attribute}
    \label{table:ablation_clustering_results_class}
    \centering
    \resizebox{\textwidth}{!}{
    \begin{tabular}{c | c c | c c}
        \toprule
         \multirow{2}{*}{Method} & \multicolumn{2}{c}{Chapman} & \multicolumn{2}{c}{PTB-XL} \\
         & Acc & AMI & Acc & AMI \\ 
        \midrule

         $\mathcal{L}_{NCE-hard}$ & $86.8$ {\scriptsize(0.7)} & $67.5$ {\scriptsize(1.1)} & $66.5$ {\scriptsize(0.1)} & $35.0$ {\scriptsize(0.0)} \\
         
         \midrule
         \multicolumn{5}{l}{$\mathcal{L}_{NCE-soft}$} \\
         \midrule
         
         $\tau_{\omega} = \infty$ & $87.3$ {\scriptsize(0.5)} & $68.2$ {\scriptsize(0.6)}  & $76.3$ {\scriptsize(0.5)} & $36.1$ {\scriptsize(1.0)}  \\
         
         $\tau_{\omega} \neq \infty$ & $89.8$ {\scriptsize(1.7)} & $72.1$ {\scriptsize(2.8)}  & $76.1$ {\scriptsize(0.2)} & $36.0$ {\scriptsize(0.4)} \\
         
         $+ \mathcal{L}_{reg}$ & $\boldsymbol{90.3}$ {\scriptsize(0.8)} & $\boldsymbol{72.8}$ {\scriptsize(1.6)} & $76.0$ {\scriptsize(0.3)} & $35.9$ {\scriptsize(0.4)} \\
         
         \bottomrule
    \end{tabular}
    }
    \end{subtable}
    ~
    \begin{subtable}{0.45\textwidth}
    \centering
    \caption{Sex and age attributes}
    \label{table:ablation_clustering_results_nonclass}
    \resizebox{\textwidth}{!}{
    \begin{tabular}{c | c c | c c}
        \toprule
         \multirow{2}{*}{Method} & \multicolumn{2}{c}{Chapman} & \multicolumn{2}{c}{PTB-XL} \\
         & sex & age & sex & age \\ 
         
         \midrule
         $\mathcal{L}_{NCE-hard}$ & $56.9$ {\scriptsize(0.2)} & $26.2$ {\scriptsize(0.0)} & $76.3$ {\scriptsize(0.7)} & $19.8$ {\scriptsize(0.0)} \\ 
         
         \midrule
         \multicolumn{5}{l}{$\mathcal{L}_{NCE-soft}$} \\
         \midrule  
         
         $\tau_{\omega} = \infty$ & $55.2$ {\scriptsize(0.5)} & $34.7$ {\scriptsize(0.3)} & $50.4$ {\scriptsize(0.1)} & $20.8$ {\scriptsize(1.0)} \\
         
         $\tau_{\omega} \neq \infty$ & $56.8$ {\scriptsize(1.8)} & $37.4$ {\scriptsize(1.0)}  & $74.3$ {\scriptsize(0.0)} & $19.2$ {\scriptsize(0.9)} \\
         
         $+ \mathcal{L}_{reg}$ & $\boldsymbol{57.4}$ {\scriptsize(1.2)} & $\boldsymbol{38.0}$ {\scriptsize(0.8)} & $73.5$ {\scriptsize(0.6)} & $19.5$ {\scriptsize(0.2)} \\

         \bottomrule
    \end{tabular}
    }
    \end{subtable}
\end{table}

\subsection{Investigating the marginal impact of design choices}
\label{sec:marginal_impact}

We have shown that CROCS reliably allows for both clustering and retrieval. In this section, we conduct several ablation studies to better understand the root cause of this reliability (see Table~\ref{table:ablation_clustering_results}). We find that, on average, the soft assignment of representations to prototypes is preferable to the hard assignment. For example, on PTB-XL, $\mathcal{L}_{NCE-soft}$ and $\mathcal{L}_{NCE-hard}$ achieve $\mathrm{Acc(class)}\approx 76.0$ and $66.5\%$, respectively. We also find that our full framework ($\mathcal{L}_{NCE-soft} + \mathcal{L}_{reg}$) performs better than, or on par with, other variants. For example, on Chapman, $\mathcal{L}_{NCE-hard}$ and $\mathcal{L}_{NCE-soft}$ $\tau_{\omega} = \infty$, and $\tau_{\omega} \neq \infty$ achieve $\mathrm{AMI(class)} = 67.5$, $68.2$, and $72.1\%$, respectively, whereas $\mathcal{L}_{NCE-soft} + \mathcal{L}_{reg}$ achieves $\mathrm{AMI(class)} = 72.8\%$. This is a positive outcome given that the regularization term's main purpose was simply to improve the interpretability of prototypes by allowing them to capture the semantic relationships between attributes. These findings extend to the retrieval setting (Appendix~\ref{appendix:marginal_impact}). 

\section{Discussion}
\label{sec:discussion}

In this paper, we proposed a supervised contrastive learning framework, entitled CROCS, for the clustering and retrieval of cardiac signals based on multiple patient attributes. In the process, we attracted representations associated with a set of attributes to learnable embeddings, termed clinical prototypes, that share such attributes and repelled them from prototypes with different attributes. We showed that CROCS outperforms the state-of-the-art method, DTC, when clustering, and retrieves relevant cardiac signals while lending itself to a higher degree of interpretability. 

We acknowledge several limitations that can be addressed in future work. We made the design choice of discretizing patient attributes (e.g., sex and age). However, these attributes, and indeed other clinical parameters, can be continuous. Therefore, extending clinical prototypes to capture this continuity could offer researchers finer control over the clustering and retrieval process. In doing so, we envision two main challenges. First, continuous attribute values imply that the number of prototypes to be learned will grow. This may pose a computational bottleneck. To alleviate this bottleneck, researchers may benefit from advancements in the field of NLP and textual representation learning. Second, to learn useful prototypes in this continuous setting, one may require sufficient labelled data associated with each attribute combination. In practice, this may be difficult to achieve, particularly for under-represented attribute groups. Furthermore, in this work, we were limited to three discrete patient attributes. Such a decision, although primarily motivated by the availability of patient meta-data alongside ECG signals, does not exploit the diverse and abundant meta-data typically available within healthcare.

\section*{Acknowledgements}

We thank the anonymous reviewers and Antong Chen for their insightful feedback. We also thank Nagat Al-Saghira and Mohammed Abdel Wahab for lending us their voice. David Clifton was supported by the EPSRC under Grants EP/P009824/1and EP/N020774/1, and by the National Institute for Health Research (NIHR) Oxford Biomedical Research Centre (BRC). The views expressed are those of the authors and not necessarily those of the NHS, the NIHR or the Department of Health. Tingting Zhu was supported by the Engineering for Development Research Fellowship provided by the Royal Academy of Engineering. 

\bibliographystyle{unsrt}
\bibliography{neurips_2021}

\section*{Checklist}

\begin{enumerate}

\item For all authors...
\begin{enumerate}
  \item Do the main claims made in the abstract and introduction accurately reflect the paper's contributions and scope?
    \answerYes{In the abstract and introduction, we claimed to have designed a supervised contrastive learning framework for the clustering and retrieval of cardiac signals. We also claimed to outperform several baseline methods. Results substantiating these claims are shown in Secs~\ref{sec:clustering} and \ref{sec:retrieval}.}
  \item Did you describe the limitations of your work?
    \answerYes{We discussed limitations around using discrete attribute values and the scalability of our framework to many attributes (see Sec.~\ref{sec:discussion}).}
  \item Did you discuss any potential negative societal impacts of your work?
    \answerNA{}
  \item Have you read the ethics review guidelines and ensured that your paper conforms to them?
    \answerYes{}
\end{enumerate}

\item If you are including theoretical results...
\begin{enumerate}
  \item Did you state the full set of assumptions of all theoretical results?
    \answerNA{}
	\item Did you include complete proofs of all theoretical results?
    \answerNA{}
\end{enumerate}

\item If you ran experiments...
\begin{enumerate}
  \item Did you include the code, data, and instructions needed to reproduce the main experimental results (either in the supplemental material or as a URL)?
    \answerYes{The supplementary material contains in-depth implementation details to reproduce our experiments. Moreover, all datasets are publicly-available. Code is currently undergoing a patent examination and can be released on a case-by-case basis.}
  \item Did you specify all the training details (e.g., data splits, hyperparameters, how they were chosen)?
    \answerYes{Details about the hyperparameters are provided in Sec.~\ref{sec:design}. Further implementation details are provided in the supplementary material.}
	\item Did you report error bars (e.g., with respect to the random seed after running experiments multiple times)?
    \answerYes{We always report our results with a standard deviation across five random seeds, e.g., Table 1, 2, and 3 in Secs.~\ref{sec:clustering}, \ref{sec:retrieval}, and \ref{sec:marginal_impact}, respectively.}
	\item Did you include the total amount of compute and the type of resources used (e.g., type of GPUs, internal cluster, or cloud provider)?
    \answerYes{These details can be found in Appendix~\ref{appendix:network}.}
\end{enumerate}

\item If you are using existing assets (e.g., code, data, models) or curating/releasing new assets...
\begin{enumerate}
  \item If your work uses existing assets, did you cite the creators?
    \answerYes{We use, and appropriately cite, two publicly-available electrocardiogram datasets (see Sec.~\ref{sec:design}).}
  \item Did you mention the license of the assets?
    \answerNA{}
  \item Did you include any new assets either in the supplemental material or as a URL?
    \answerNA{}
  \item Did you discuss whether and how consent was obtained from people whose data you're using/curating?
    \answerNA{}
  \item Did you discuss whether the data you are using/curating contains personally identifiable information or offensive content?
    \answerNA{}
\end{enumerate}

\item If you used crowdsourcing or conducted research with human subjects...
\begin{enumerate}
  \item Did you include the full text of instructions given to participants and screenshots, if applicable?
    \answerNA{}
  \item Did you describe any potential participant risks, with links to Institutional Review Board (IRB) approvals, if applicable?
    \answerNA{}
  \item Did you include the estimated hourly wage paid to participants and the total amount spent on participant compensation?
    \answerNA{}
\end{enumerate}

\end{enumerate}


\clearpage

\appendix

\section{Datasets}
\label{appendix:datasets}

\subsection{Data pre-processing}
\label{appendix:data_description}

For all of the datasets, frames consisted of 2500 samples and consecutive frames had no overlap with one another. Data splits were always performed at the patient-level.

\textbf{Chapman} \cite{Zheng2020}. Each ECG recording was originally 10 seconds with a sampling rate of 500Hz. We downsample the recording to 250Hz and therefore each ECG frame in our setup consisted of 2500 samples. We follow the labelling setup suggested by \cite{Zheng2020} which resulted in four classes: Atrial Fibrillation, GSVT, Sudden Bradychardia, Sinus Rhythm. The ECG frames were normalized in amplitude between the values of 0 and 1. 

\textbf{PTB-XL} \cite{Wagner2020}. Each ECG recording was originally 10 seconds with a sampling rate of 500Hz. We extract 5-second non-overlapping segments of each recording generating frames of length 2500 samples. We follow the diagnostic class labelling setup suggested by \cite{Strodthoff2020} which resulted in five classes: Conduction Disturbance (CD), Hypertrophy (HYP), Myocardial Infarction (MI), Normal (NORM), and Ischemic ST-T Changes (STTC). Furthermore, we only consider ECG segments with one label assigned to them. The ECG frames were standardized to follow a standard Gaussian distribution. 

\subsection{Data samples}
\label{appendix:instances}

In this section, we outline the number of instances used during training, validation, and testing for the Chapman and PTB-XL datasets (see Table~\ref{table:data_splits}).

\begin{table}[h]
\small
\centering
\caption{Number of instances (number of patients) used during training. These represent sample sizes for all 12 leads.}
\vskip 0.1in
\label{table:data_splits}
\begin{tabular}{c | c c c}
\toprule
Dataset & Train & Validation & Test\\
\midrule
\multirow{1}{*}{Chapman}&76,614 (6,387)&25,524 (2,129)&25,558 (2,130)\\
\multirow{1}{*}{PTB-XL}& 22,670 (11,335) &3,284 (1,642)&3,304 (1,152)\\
\bottomrule
\end{tabular}
\end{table}

\clearpage

\section{Implementation details}
\label{appendix:network}

In this section, we outline the neural network architectures used for our experiments. More specifically, we use the architecture shown in Table~\ref{table:network_architecture} for all experiments pertaining to the Chapman dataset. Given the size of the PTB-XL dataset and the relative complexity of the corresponding task (at least at the disease class level), we opted for a more complex network. We modified the ResNet18 architecture whereby the number of blocks per layer was reduced from two to one, effectively reducing the number of parameters by a factor of two. We chose this architecture after experimenting with several variants. Experiments were conducted using PyTorch \cite{Paszke2019} and an NVIDIA Quadro RTX 6000 GPU. Each  training and validation epoch took approximately 2 minutes, and 20 seconds to complete, respectively. 

\begin{table}[h]
\small
\centering
\caption{Network architecture used for experiments conducted on the Chapman dataset. \textit{K}, \textit{C}\textsubscript{in}, and \textit{C}\textsubscript{out} represent the kernel size, number of input channels, and number of output channels, respectively. A stride of 3 was used for all convolutional layers. $E$ represents the dimension of the final representation.}
\vskip 0.1in
\label{table:network_architecture}
\begin{tabular}{c c c}
\toprule
Layer Number &Layer Components&Kernel Dimension\\
\midrule
	\multirow{5}{*}{1}&Conv 1D & 7 x 1 x 4 (\textit{K} x \textit{C}\textsubscript{in} x \textit{C}\textsubscript{out})\\
										& BatchNorm &\\
										& ReLU& \\
										& MaxPool(2)& \\
										& Dropout(0.1) &\\
	\midrule
	\multirow{5}{*}{2}&Conv 1D & 7 x 4 x 16\\
										& BatchNorm& \\
										& ReLU &\\
										& MaxPool(2) &\\
										& Dropout(0.1)& \\
	\midrule
	\multirow{5}{*}{3}&Conv 1D & 7 x 16 x 32 \\
										& BatchNorm &\\
										& ReLU &\\
										& MaxPool(2) &\\
										& Dropout(0.1) &\\
	\midrule
	\multirow{2}{*}{4}&Linear&320 x $E$ \\
										& ReLU &\\
\bottomrule
\end{tabular}
\end{table}

\begin{table}[h]
\small
\centering
\caption{Batchsize and learning rates used for training with different datasets. The Adam optimizer was used for all experiments.}
\label{table:batchsize}
\begin{tabular}{c | c c }
\toprule
Dataset & Batchsize & Learning Rate\\
\midrule 
Chapman & 256 & 10\textsuperscript{-4}\\
PTB-XL & 128 & 10\textsuperscript{-5}\\
\bottomrule
\end{tabular}
\end{table}

\clearpage

\section{Baseline implementations}
\label{appendix:baseline_methods}

\paragraph{DeepCluster}
In the implementation by Caron \textit{et al.} \cite{Caron2018}, a forward pass of each instance in the training set is performed. This generates a set of representation which are then clustered, in an unsupervised manner, using $k$-means. This involves a decision regarding the value of K, i.e., the number of clusters. In our supervised setting, we have this information available and therefore set the value of K to be equal to the number of distinct cardiac arrhythmia classes. Once the clustering is complete, each instance is assigned a pseudo-label according to the cluster to which it belongs. Such pseudo-labels are used as the ground-truth for supervised training during the next epoch. We repeat this process after each epoch for a total of 30 epochs after realizing that the validation loss plateaus at that point.

\paragraph{IIC}
In this implementation, the network is tasked with maximizing the mutual information between the representation of an instance and that of its perturbed counterpart. Such perturbations must be class-preserving and, in computer vision, consist of random crops, rotations, and modifications to the brightness of the images. In our setup involving time-series data, we perturb instances by using additive Gaussian noise in order to avoid erroneously flipping the class of a particular instance. In addition to the aforementioned, we implement the auxiliary over-clustering method suggested by the authors. This approach allows one to model additional 'distractor' classes that may be present in the dataset, and was shown by \cite{Ji2019} to improve generalization performance. In our setup, we set the number of total clusters to the number of attribute combinations, $M$. 

\paragraph{SeLA}
In this implementation, each instance is \textit{assigned} a posterior probability distribution. For all instances, this results in an assigned matrix of posterior probability distributions. Each instance's label is obtained by identifying the index associated with the largest posterior probability distribution. Deriving the aforementioned matrix is the crux of SeLA. It does by solving the Sinkhorn-Knopp algorithm under the assumption that the dataset can be evenly split into K clusters. Our setup does not deviate from the original implementation found in \cite{Asano2020}.  

\paragraph{DeepTransferCluster} In this implementation, the distance between each representation and each cluster prototype is calculated to generate a probability distribution over classes, $p$. The distribution, $p$, is encouraged to be similar to a target distribution, $z$, by minimizing the KL divergence of these two distributions. In the original unsupervised implementation, the target distribution is a sharper version of the empirical distribution \cite{Han2019}. In our supervised implementation, we initialize the prototypes similarly to our approach and modify the target distribution to incorporate labels. As with our soft-assignment, we aim for a target distribution that reflects discrepancies, $d$, between the representation attributes, $A_{i}$, and the prototype attributes, $A_{j}$. Mathematically, our target distribution, $z$, is as follows:
\begin{equation}
    z_{j} = \frac{e^{\omega_{ij}}}{\sum_{l}^{|L|} e^{\omega_{il}}}
\end{equation}
\begin{equation}
    \label{eq:weights}
    \omega_{ij} = \begin{cases} \frac{e^{d(A_{i},A_{j})}}{\sum_{l}^{|L|} e^{d(A_{i},A_{l})}}& \mbox{if   }     \alpha_{1}^{i}=\alpha_{1}^{j} \\
    0 & \mbox{otherwise}
    \end{cases}
\end{equation}
\begin{equation}
    \label{eq:discrepancy}
    d(A_{i},A_{j})  = \frac{1}{\tau_{\omega}} \cdot [ \delta(\alpha_{c}^{i}=\alpha_{c}^{j}) + 
     \delta(\alpha_{s}^{i}=\alpha_{s}^{j}) + 
     \delta(\alpha_{a}^{i}=\alpha_{a}^{j}) ]
\end{equation}
\paragraph{K-means EP} In this implementation by Gee \textit{et al.} \cite{Gee2019}, each instance is first passed through the encoder network to generate a representation. This representation serves multiple functions: a) it is passed through the decoder network to reconstruct the input, and b) passed through a prototype network that works as follows. The Euclidean distance between the representation and $M$ randomly-initialized embeddings (prototypes) is calculated to generate a single $M$-dimensional representation. This newly-generated representation is then passed through a linear classification head to predict the cardiac arrhythmia class associated with the original instance. In our setup, we set the number of prototypes to coincide with the number of clinical prototypes that we use. For clustering, we apply the $k$-means algorithm to the representations learned via this framework. 

\paragraph{Deep Temporal Clustering Representation} In this implementation by Ma \textit{et al.} \cite{Ma2019}, the network consists of three main components: 1) an encoder, 2) a decoder, and 3) a classifier head. A synthetic version of each instance is first generated by permuting a certain fraction, $\alpha$, of the time-points in the original instance. The original instance and its synthetic counterpart are then passed through the encoder to obtain a pair of representations (a real and synthetic one). The classifier is tasked with identifying whether such representations are real or fake (binary classification akin to discriminator in generative adversarial networks). Moreover, the decoder reconstructs the original instance by operating on the real representation. Lastly, the $k$-means loss is approximated based on the Gram matrix of the mini-batch of real representations. We follow the original implementation, and choose $\alpha=0.2$, and $\lambda=10^{-3}$ as the coefficient of the $k$-means loss in the objective function. 

\section{Effect of embedding dimension, $E$, on clustering}
\label{appendix:effect_of_embedding_size}

\begin{figure}[!h]
    \centering
    \begin{subfigure}{0.45\textwidth}
        \includegraphics[width=1\textwidth]{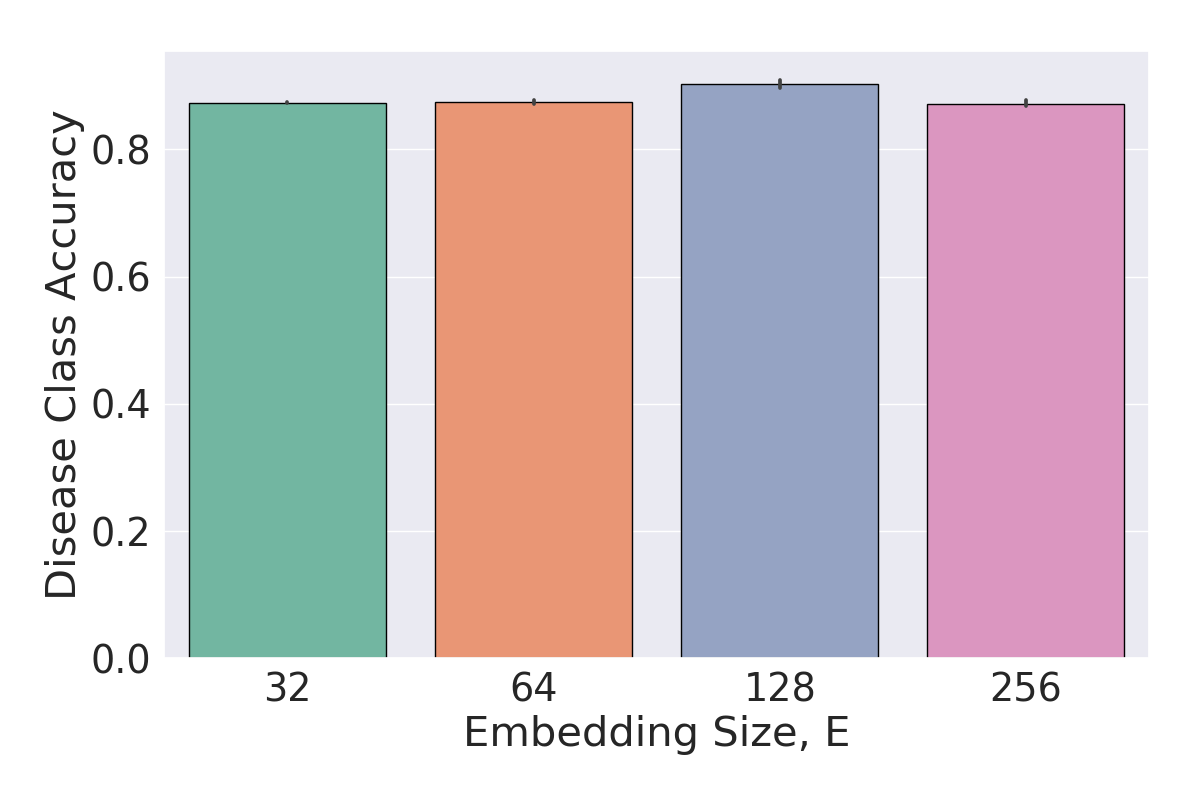}
        \caption{}
        \label{fig:effect_of_embedding_size_class}    
    \end{subfigure}
    ~
    \begin{subfigure}{0.45\textwidth}
        \includegraphics[width=1\textwidth]{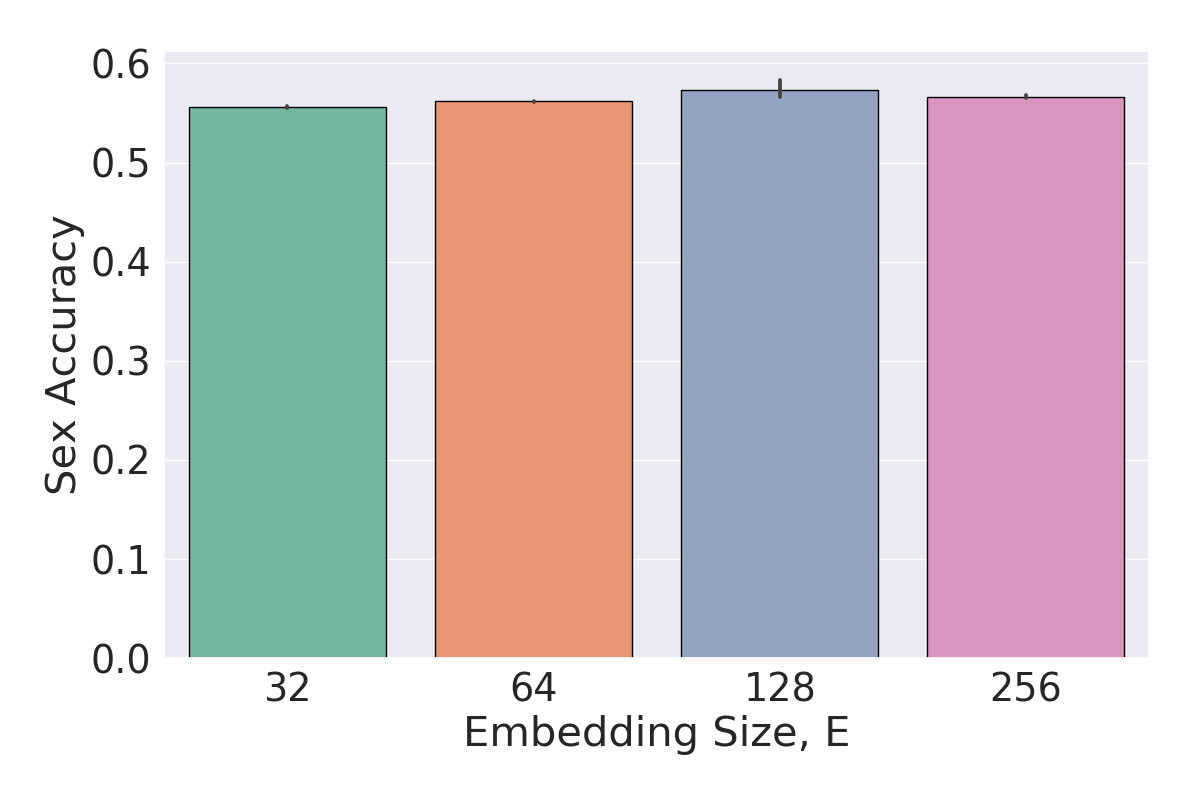}
        \caption{}
        \label{fig:effect_of_embedding_size_sex}    
    \end{subfigure}
    ~
    \begin{subfigure}{0.45\textwidth}
        \includegraphics[width=1\textwidth]{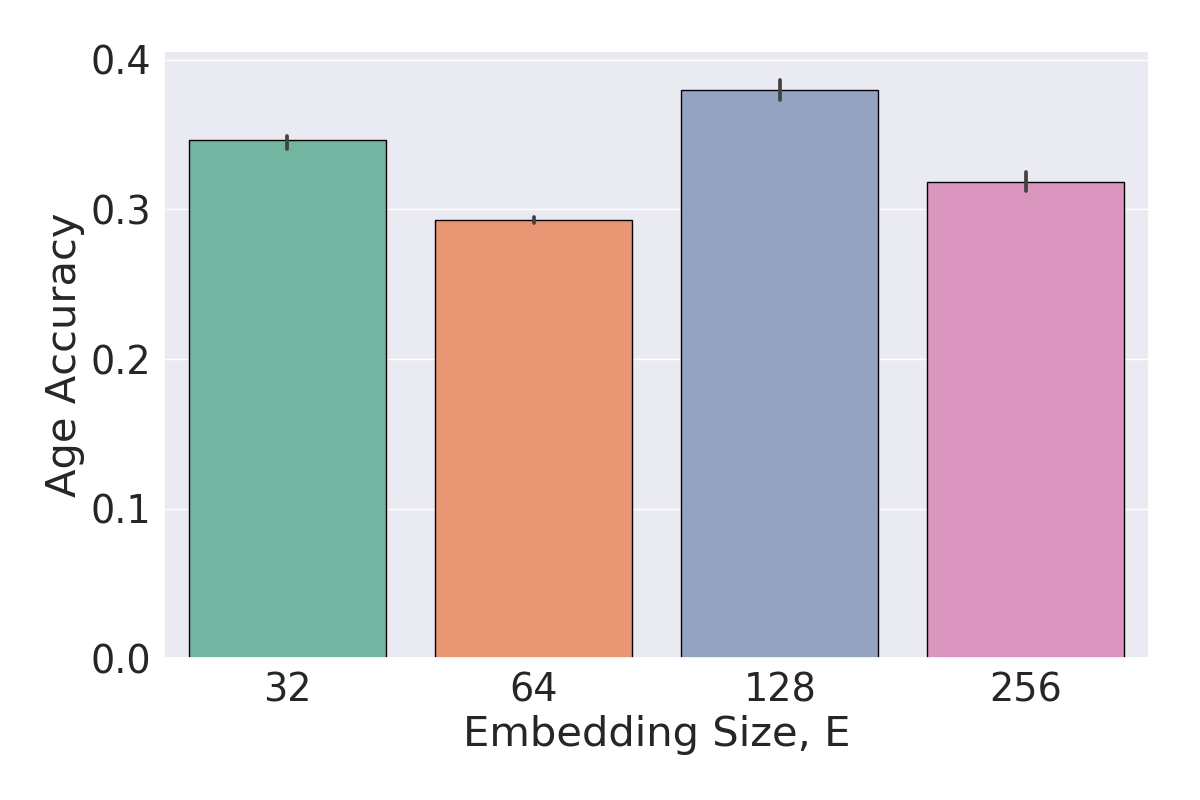}
        \caption{}
        \label{fig:effect_of_embedding_size_age}    
    \end{subfigure}
    \caption{\textbf{Effect of embedding dimension, $E$, on the clustering performance of our framework.} Results are shown for the (a) disease class, (b) sex, and (c) age attributes across five random seeds. The error bars represent one standard deviation from the mean. We find that the embedding dimension has a minimal effect on the performance when evaluating based on disease class and sex attributes. An effect is more pronounced when clustering based on the age attribute.}
    \label{fig:effect_of_embedding_size}    
\end{figure}

In this section, we explore the effect of the embedding dimension, $E$, on the clustering performance of our framework. Specifically, we experiment with $E \in \{32,64,128,256\}$ on the Chapman dataset and present the accuracy of the attribute assignments (disease class, sex, and age) in Fig.~\ref{fig:effect_of_embedding_size}. We find that the embedding dimension has minimal impact on the clustering performance of our framework when evaluated on the disease class and sex patient attributes. This is evident in Fig.~\ref{fig:effect_of_embedding_size_class} where the $\mathrm{Acc(class)} \approx 0.85$ across all embedding dimensions and in Fig.~\ref{fig:effect_of_embedding_size_sex} where the $\mathrm{Acc(class)} \approx 0.55$ across all embedding dimensions. We do, however, find that an embedding dimension, $E=128$, is favourable when evaluating the clustering performance based on the patient age assignments. This can be seen in Fig.~\ref{fig:effect_of_embedding_size_age} where $\mathrm{Acc(class)} \approx 0.38$ at $E=128$, whereas $\mathrm{Acc(class)} < 0.34$ for the remaining embedding dimensions.

\clearpage

\section{Effect of $\beta$ on clustering and retrieval}
\label{appendix:effect_of_beta}

In this section, we examine the effect of $\beta$, as used in the regularization term (Eq.~\ref{eq:full_loss}), on the clustering and retrieval performance of our framework. We conduct the same clustering and retrieval experiments as those found in the main manuscript and experiment with $\beta=[0.05,0.1,0.2,0.4]$. The results of these experiments are presented in Tables~\ref{table:beta_clustering_results} and \ref{table:beta_retrieval_results}. In both settings, we find that $\beta=0.2$ is preferable to the remaining values of $\beta$. This is evident by the higher clustering and retrieval performance. For example, at $\beta=0.2$, $\mathrm{Acc(class)}=76.0$ whereas at the remaining $\beta$ values, $\mathrm{Acc(class)} \approx 66.0$, reflecting a difference of $11\%$. Furthermore, in the retrieval setting, with $K=1$ and the \# attribute matches $\geq 1$, the precision at $\beta=0.2$ is $68.8$ whereas at the remaining values of $\beta$, the precision $< 67.5$. These findings are consistent with our expectations given that $\beta$ controls the distance between clinical prototypes, which, in turn, impacts their utility as centroids for clustering and as queries for retrieval.  

\begin{table}[!h]
    \centering
    \small
    \caption{Effect of $\beta$ on the clustering performance on representations in the validation set of PTB-XL. Evaluation is based on (a) class attribute and (b) sex and age attributes. Results are averaged across five random seeds. Brackets indicate standard deviation and bold reflects the top-performing $\beta$ value.}
    \label{table:beta_clustering_results}
    \begin{subtable}{0.7\textwidth}
    \caption{Cardiac arrhythmia class attribute}
    \label{table:beta_clustering_results_class}
    \centering
    \begin{tabular}{c | c c}
        \toprule
         \multirow{2}{*}{$\beta$} & \multicolumn{2}{c}{PTB-XL} \\
         & Acc & AMI \\ 
        \midrule

         0.05 & $66.0$ {\scriptsize(0.3)} & $33.3$ {\scriptsize(0.0)} \\
         0.1 & $65.7$ {\scriptsize(0.0)} & $33.0$ {\scriptsize(0.6)} \\
         0.2 & $\boldsymbol{76.0}$ {\scriptsize(0.3)} & $\boldsymbol{35.9}$ {\scriptsize(0.4)} \\
         0.4 & $65.9$ {\scriptsize(0.6)} & $34.7$ {\scriptsize(0.2)} \\

         \bottomrule
    \end{tabular}
    \end{subtable}
    ~
    \begin{subtable}{0.7\textwidth}
    \centering
    \vskip 0.1in
    \caption{Sex and age attributes}
    \label{table:beta_clustering_results_nonclass}
    \begin{tabular}{c | c c}
        \toprule
         \multirow{2}{*}{Method} & \multicolumn{2}{c}{PTB-XL} \\
         & Sex & Age \\ 
         \midrule
         
        0.05 & $72.2$ {\scriptsize(0.2)} & $32.4$ {\scriptsize(0.0)} \\
        0.1 & $72.9$ {\scriptsize(0.8)} & $31.6$ {\scriptsize(0.5)} \\
        0.2 & $\boldsymbol{73.5}$ {\scriptsize(0.6)} & $19.5$ {\scriptsize(0.2)}\\
        0.4 & $72.5$ {\scriptsize(0.8)} & $32.8$ {\scriptsize(0.2)} \\

         \bottomrule
    \end{tabular}
    \end{subtable}
\end{table}

\begin{table}[!h]
\small
\centering
\caption{Effect of $\beta$ on precision of CROCS when retrieving the closest $K$ representations from the validation set of PTB-XL. Results shown are based on the number of attributes shared between the prototypes and the retrieved cardiac signals, and are averaged across five random seeds. Brackets indicate standard deviation and bold reflects the top-performing $\beta$ value.}
    \label{table:beta_retrieval_results}
    \centering
    \begin{tabular}{c c | c c c}
        \toprule
        \multirow{2}{*}{\# attribute matches} & \multirow{2}{*}{$\beta$} & \multicolumn{3}{c}{PTB-XL} \\
        & & $K=1$ & $5$ & $10$ \\
        \midrule
        
        \multirow{4}{*}{$\geq1$} & 0.05 & $63.5$ {\scriptsize(0.0)} & $100.0$ {\scriptsize(0.0)} & $100.0$ {\scriptsize(0.0)}  \\
        
        & 0.1 & $63.5$ {\scriptsize(7.0)} & $100.0$ {\scriptsize(0.0)} & $100.0$ {\scriptsize(0.0)} \\
        
        & 0.2 & $\boldsymbol{68.8}$ {\scriptsize(8.1)} & $100.0$ {\scriptsize(0.0)} & $100.0$ {\scriptsize(0.0)}  \\
        
        & 0.4 & $67.5$ {\scriptsize(7.7)} & $97.0$ {\scriptsize(2.4)} & $100.0$ {\scriptsize(0.0)} \\
        
        \midrule 
        
        \multirow{4}{*}{$\geq2$} & 0.05 & $18.0$ {\scriptsize(4.5)} & $55.0$ {\scriptsize(5.0)} & $69.0$ {\scriptsize(9.3)} \\
        
        & 0.1 & $15.5$ {\scriptsize(1.0)} & $49.5$ {\scriptsize(4.0)} & $70.5$ {\scriptsize(6.2)}  \\
        
        & 0.2 & $\boldsymbol{19.4}$ {\scriptsize(6.4)} & $\boldsymbol{78.8}$ {\scriptsize(11.9)} & $\boldsymbol{93.8}$ {\scriptsize(4.4)} \\
        
        & 0.4 & $13.0$ {\scriptsize(4.0)} & $50.0$ {\scriptsize(1.6)} & $80.5$ {\scriptsize(1.0)} \\
        
        \midrule
        
        \multirow{4}{*}{$=3$} & 0.05 & $1.5$ {\scriptsize(2.0)} & $6.0$ {\scriptsize(2.0)} & $10.0$ {\scriptsize(0.0)} \\

        & 0.1 & $0.5$ {\scriptsize(1.0)} & $5.5$ {\scriptsize(1.0)} & $8.0$ {\scriptsize(1.0)} \\

        & 0.2 &  $1.3$ {\scriptsize(2.5)} & $\boldsymbol{14.4}$ {\scriptsize(7.0)} & $\boldsymbol{26.3}$ {\scriptsize(8.0)} \\
        
        & 0.4 & $0.5$ {\scriptsize(1.0)} & $2.5$ {\scriptsize(0.0)} & $10.0$ {\scriptsize(1.6)} \\
        
        \bottomrule
\end{tabular}
\end{table}

\clearpage

\section{Performance of CROCS with Less Labelled Data}
\label{appendix:effect_of_data}

In this section, we explore the effect of less labelled data on the performance of CROCS. More precisely, we reduce the amount of labelled training data 10-fold while keeping the amount of unlabelled data fixed. In Tables~\ref{table:effect_of_data_clustering_results} and \ref{table:effect_of_data_retrieval_results}, we present the results of these experiments in the clustering and retrieval settings, respectively.

\begin{table}[!h]
    \centering
    \small
    \caption{\textbf{Clustering performance on the validation set of Chapman and PTB-XL when CROCS is trained with 10\% of labelled data.} Evaluation is based on (a) class and (b) sex and age attributes. Results are averaged across five random seeds. Brackets indicate standard deviation and bold reflects the top-performing method. The asterisk ($*$) indicates that the algorithm could not be solved. We show that $\mathrm{CP \ CROCS}$ outperforms the remaining methods regardless of patient attribute.}
    \label{table:effect_of_data_clustering_results}
    \begin{subtable}{0.48\textwidth}
    \caption{Cardiac arrhythmia class attribute}
    \label{table:clustering_results_class}
    \centering
    \resizebox{\textwidth}{!}{
    \begin{tabular}{c | c c | c c}
        \toprule
         \multirow{2}{*}{Method} & \multicolumn{2}{c}{Chapman} & \multicolumn{2}{c}{PTB-XL} \\
         & Acc & AMI & Acc & AMI \\ 
        \midrule

         SeLA \cite{Asano2020} & $*$ & $*$ & $0.10$ {\scriptsize(0.0)} & $0.02$ {\scriptsize(0.0)} \\ 
         
         DC \cite{Caron2018} & $21.0$ {\scriptsize(0.0)} & $0.5$ {\scriptsize(0.6)} & $10.5$ {\scriptsize(0.0)} & $4.3$ {\scriptsize(0.9)} \\
         
         IIC \cite{Ji2019} & $27.2$ {\scriptsize(0.3)} & $0.6$ {\scriptsize(0.01)} & $24.3$ {\scriptsize(2.8)} & $0.4$ {\scriptsize(0.7)} \\
        
         DTCR \cite{Ma2019} & $34.3$ {\scriptsize(0.9)} & $3.3$ {\scriptsize(0.0)} & 24.1 {\scriptsize(0.5)} & $0.8$ {\scriptsize(0.3)} \\
        
         DTC \cite{Han2019} & $46.3$ {\scriptsize(2.6)} & $11.8$ {\scriptsize(2.2)} & $48.4$ {\scriptsize(3.9)} & $0.3$ {\scriptsize(0.5)} \\
         
         KM raw & $28.4$ {\scriptsize(1.2)} & $0.3$ {\scriptsize(0.0)} & - & - \\
         
         KM EP \cite{Gee2019} & $64.9$ {\scriptsize(4.7)} & $44.3$ {\scriptsize(3.4)} & $45.1$ {\scriptsize(1.9)} & $20.7$ {\scriptsize(1.3)} \\
         
         KM CROCS & $71.1$ {\scriptsize(4.6)} & $52.7$ {\scriptsize(2.3)} & $47.7$ {\scriptsize(3.4)} & $20.1$ {\scriptsize(0.9)} \\
         
         TP CROCS & $75.5$ {\scriptsize(0.2)} & $56.8$ {\scriptsize(0.5)} & $47.3$ {\scriptsize(2.8)} & $19.9$ {\scriptsize(1.4)} \\

         CP CROCS & $\boldsymbol{82.7}$ {\scriptsize(0.4)} & $\boldsymbol{61.8}$ {\scriptsize(0.8)} & $\boldsymbol{71.4}$ {\scriptsize(0.0)} & $\boldsymbol{28.8}$ {\scriptsize(0.8)} \\
         \bottomrule
    \end{tabular}
    }
    \end{subtable}
    ~
    \begin{subtable}{0.48\textwidth}
    \centering
    \caption{Sex and age attributes}
    \label{table:clustering_results_nonclass}
    \resizebox{\textwidth}{!}{
    \begin{tabular}{c | c c | c c}
        \toprule
         \multirow{2}{*}{Method} & \multicolumn{2}{c}{Chapman} & \multicolumn{2}{c}{PTB-XL} \\
         & sex & age & sex & age \\ 
         \midrule
         
         DTCR \cite{Ma2019} & $52.2$ {\scriptsize(1.2)} & $26.7$ {\scriptsize(0.3)} & $51.7$ {\scriptsize(1.4)} & $29.3$ {\scriptsize(1.2)} \\ 
         
         DTC \cite{Han2019} & $53.1$ {\scriptsize(0.6)} & $26.8$ {\scriptsize(0.0)} & $52.2$ {\scriptsize(0.0)} & $33.5$ {\scriptsize(1.6)} \\

         KM EP \cite{Gee2019} & $56.4$ {\scriptsize(0.1)} & $30.2$ {\scriptsize(0.9)} & $51.0$ {\scriptsize(0.7)} & $37.1$ {\scriptsize(1.5)} \\

         KM CROCS & $\boldsymbol{56.2}$ {\scriptsize(0.1)} & $\boldsymbol{30.9}$ {\scriptsize(0.8)} & $51.1$ {\scriptsize(0.7)} & $36.0$ {\scriptsize(1.1)} \\
         
         TP CROCS & $53.1$ {\scriptsize(1.4)} & $29.0$ {\scriptsize(0.6)} & $60.6$ {\scriptsize(2.3)} & $36.7$ {\scriptsize(1.0)} \\

         CP CROCS & $51.0$ {\scriptsize(0.4)} & $29.7$ {\scriptsize(1.0)} & $\boldsymbol{67.5}$ {\scriptsize(1.0)} & $\boldsymbol{42.0}$ {\scriptsize(6.5)} \\
         \bottomrule
    \end{tabular}
    }
    \end{subtable}
\end{table}

\begin{table}[!h]
\small
\centering
\caption{\textbf{Precision of $\boldsymbol{K}$ retrieved representations, $\boldsymbol{v}$, in the validation set of Chapman and PTB-XL, that are closest to the query when CROCS is trained with 10\% of labelled data.} Results are shown for partial and exact matches of the attributes (\# attribute matches) represented by the query and retrieved cardiac signals, and are averaged across five random seeds. Brackets indicate standard deviation and bold reflects the top-performing method. The strong performance of $\mathrm{TP \ CROCS}$ provides evidence in support of our CROCS framework.}
    \label{table:effect_of_data_retrieval_results}
    \centering
    \vskip 0.05in
    \resizebox{0.8\textwidth}{!}{
    \begin{tabular}{c c | c c c | c c c}
        \toprule
        \# attribute & \multirow{2}{*}{Query} & \multicolumn{3}{c}{Chapman} & \multicolumn{3}{c}{PTB-XL} \\
        matches & & $K=1$ & $5$ & $10$ & $1$ & $5$ & $10$ \\
        \midrule
        
        \multirow{3}{*}{$\geq1$} & DTC \cite{Han2019} & $71.9$ {\scriptsize(0.0)} & $100.0$ {\scriptsize(0.0)} & $100.0$ {\scriptsize(0.0)} & $70.0$ {\scriptsize(0.0)} & $100.0$ {\scriptsize(0.0)} & $100.0$ {\scriptsize(0.0)} \\
        
        & TP CROCS & $\boldsymbol{91.3}$ {\scriptsize(1.3)} & $98.1$ {\scriptsize(3.8)} & $100.0$ {\scriptsize(0.0)} & $\boldsymbol{91.8}$ {\scriptsize(3.4)} & $99.5$ {\scriptsize(1.0)} & $100.0$ {\scriptsize(0.0)} \\
        
        & CP CROCS & $89.4$ {\scriptsize(2.5)} & $98.1$ {\scriptsize(3.8)} & $100.0$ {\scriptsize(0.0)} & $88.0$ {\scriptsize(6.2)} & $100.0$ {\scriptsize(0.0)} & $100.0$ {\scriptsize(0.0)} \\
        
        \midrule 
        \multirow{3}{*}{$\geq2$} & DTC \cite{Han2019} & $25.0$ {\scriptsize(0.0)} & $71.3$ {\scriptsize(1.3)} & $89.0$ {\scriptsize(4.6)} & $21.0$ {\scriptsize(0.0)} & $50.1$ {\scriptsize(5.0)} & $78.5$ {\scriptsize(8.0)}\\
        
        & TP CROCS & $\boldsymbol{53.8}$ {\scriptsize(3.6)} & $85.6$ {\scriptsize(7.0)} & $92.5$ {\scriptsize(1.5)} & $\boldsymbol{62.6}$ {\scriptsize(12.7)} & $\boldsymbol{91.3}$ {\scriptsize(4.8)} & $94.9$ {\scriptsize(1.6)} \\
        
        & CP CROCS & $\boldsymbol{51.3}$ {\scriptsize(2.5)} & $\boldsymbol{86.9}$ {\scriptsize(7.5)} & $\boldsymbol{93.8}$ {\scriptsize(7.1)} & $57.5$ {\scriptsize(5.2)} & $90.0$ {\scriptsize(5.2)} & $\boldsymbol{97.5}$ {\scriptsize(3.2)} \\
        \midrule
        \multirow{3}{*}{$=3$} & DTC \cite{Han2019} & $3.1$ {\scriptsize(0.0)} & $12.5$ {\scriptsize(0.0)} & $21.9$ {\scriptsize(2.0)} & $2.5$ {\scriptsize(0.0)} & $12.0$ {\scriptsize(1.0)} & $20.5$ {\scriptsize(4.0)}\\

        & TP CROCS & $11.3$ {\scriptsize(1.5)} & $\boldsymbol{30.0}$ {\scriptsize(3.2)} & $40.6$ {\scriptsize(4.4)} & $10.3$ {\scriptsize(3.2)} & $\boldsymbol{28.7}$ {\scriptsize(4.4)} & $37.9$ {\scriptsize(6.4)} \\

        & CP CROCS & $11.3$ {\scriptsize(1.5)} & $28.8$ {\scriptsize(7.0)} & $\boldsymbol{43.8}$ {\scriptsize(5.2)} & $9.5$ {\scriptsize(1.9)} & $27.5$ {\scriptsize(0.0)} & $\boldsymbol{39.0}$ {\scriptsize(5.2)} \\
        \bottomrule
\end{tabular}
}
\end{table}

We find that, in both the clustering and retrieval settings, our framework, CROCS, continues to generalize well and outperform the baseline methods. For example, on Chapman, CP CROCS achieves $\mathrm{Acc(class)}=0.83$, whereas KM EP, the next best baseline method, achieves $\mathrm{Acc(class)}=0.65$. This relative improvement also holds on the PTB-XL dataset. Overall, such a finding provides evidence that CROCS is relatively robust to the amount of labelled training data that are available and can thus be useful in realistic settings characterized by scarce, labelled data.

\clearpage

\section{Deploying clinical prototypes in the retrieval setting}
\label{appendix:retrieval}

\subsection{Chapman}

\begin{figure}[!h]
    \centering
    \begin{subfigure}{1\textwidth}
    \centering
	\includegraphics[width=0.48\textwidth]{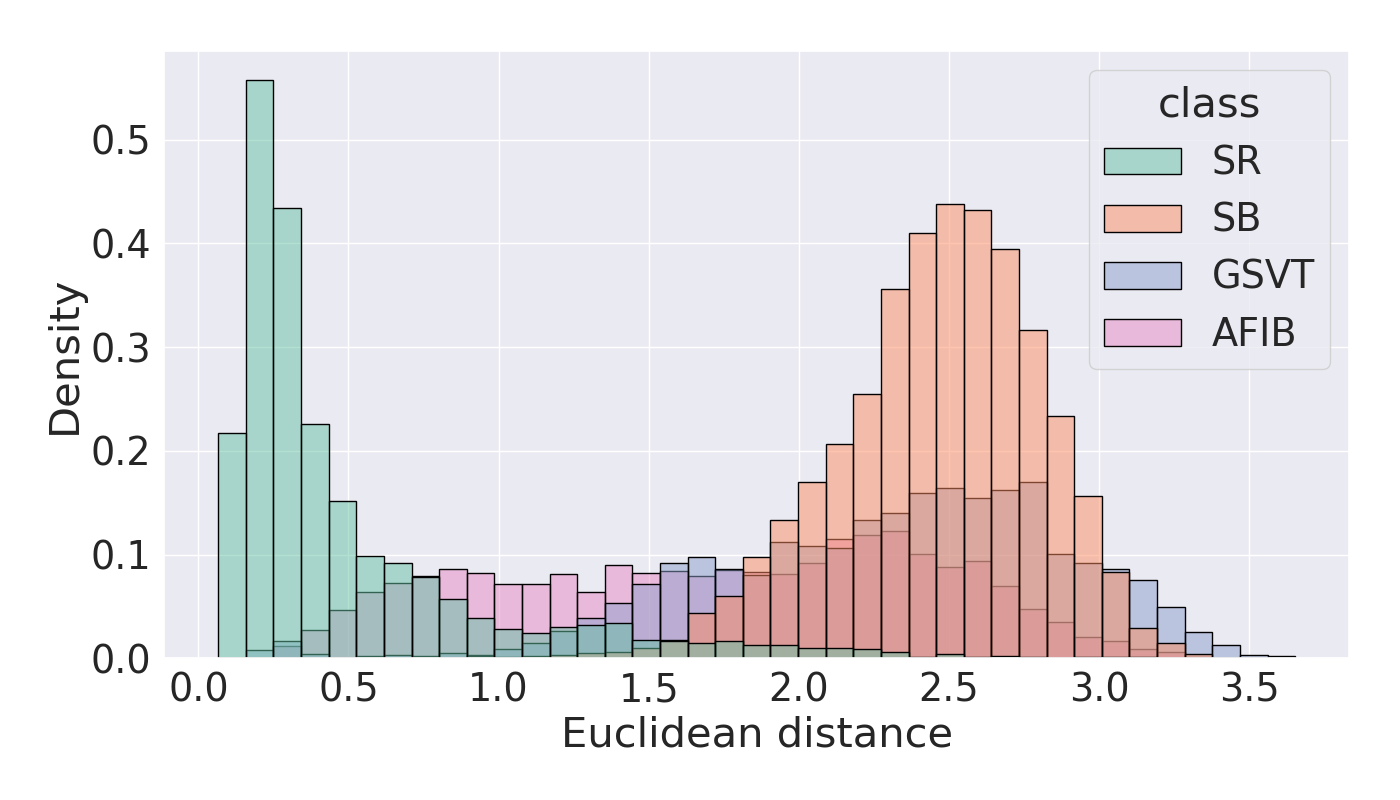}
	\end{subfigure}
	~
	\begin{subfigure}{1\textwidth}
    \centering
	\includegraphics[width=0.48\textwidth]{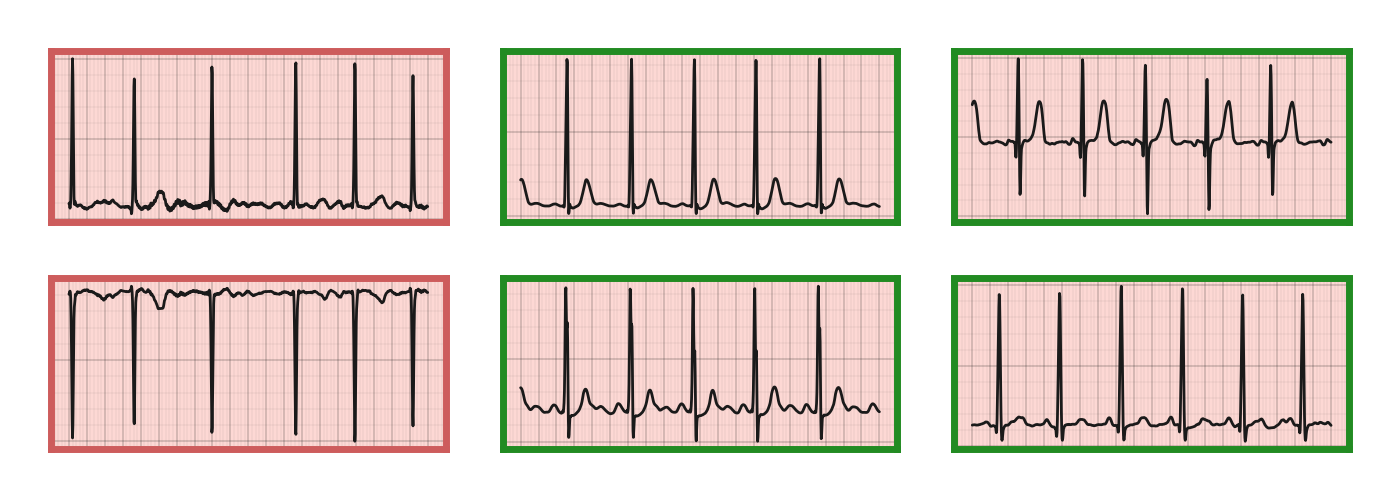}
	\caption{TP CROCS query $\{\mathrm{SR},\mathrm{male},\mathrm{under \ 49}\}$}
	\label{fig:chapman_distances_crocs}
	\end{subfigure}
	\caption{\textbf{Qualitative retrieval performance of a TP CROCS query.} (\textbf{top row}) Euclidean distance from a query to representations in the validation set of Chapman. (\textbf{bottom row}) Six closest cardiac signals to each query. The query is associated with a set of patient attributes $\{ \mathrm{disease}, \mathrm{sex}, \mathrm{age} \}$. Retrieved cardiac signals with green borders indicate those whose class attribute matches that of the query. We see that the mean representation query is closer to representations of the same class ($\mathrm{SR}$) than to those of a different class, and thus retrieves relevant cardiac signals.}
	\label{fig:chapman_crocs_top_k_ecgs}
\end{figure}

In the main manuscript, we qualitatively evaluated the retrieval performance of a DTC-derived prototype and a clinical prototype on the Chapman dataset. In this section, we continue this evaluation however for a different query; a TP CROCS query, which reflects the average of representations associated with a set of patient attributes. In Fig.~\ref{fig:chapman_crocs_top_k_ecgs} (top row), we present the distributions of the Euclidean distance between the query and the representations in the validation set of Chapman. In Fig.~\ref{fig:chapman_crocs_top_k_ecgs} (bottom row), we illustrate the six cardiac signals that are closest to the query. We find that the query is closer to representations of the class attribute than to those of a different class attribute. This is evident by the long tail of distance values exhibited between representation with $\mathrm{SR}$ and the query $\{\mathrm{SR},\mathrm{male},\mathrm{under \ 49}\}$. 

\clearpage

\subsection{PTB-XL}

\begin{figure}[!h]
    \centering
    \begin{subfigure}{0.48\textwidth}
    \centering
	\includegraphics[width=1\textwidth]{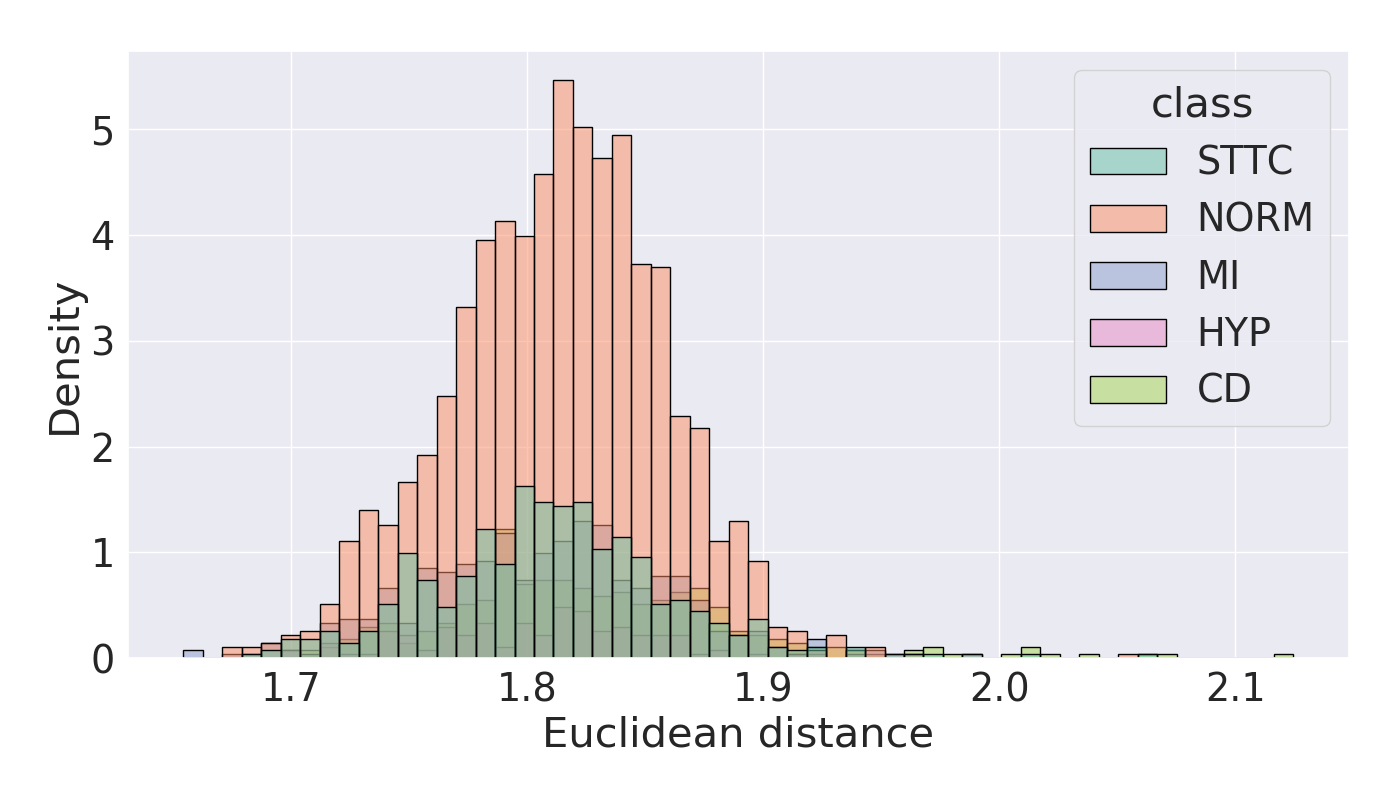}
	\end{subfigure}
	~
    \begin{subfigure}{0.48\textwidth}
    \centering
	\includegraphics[width=1\textwidth]{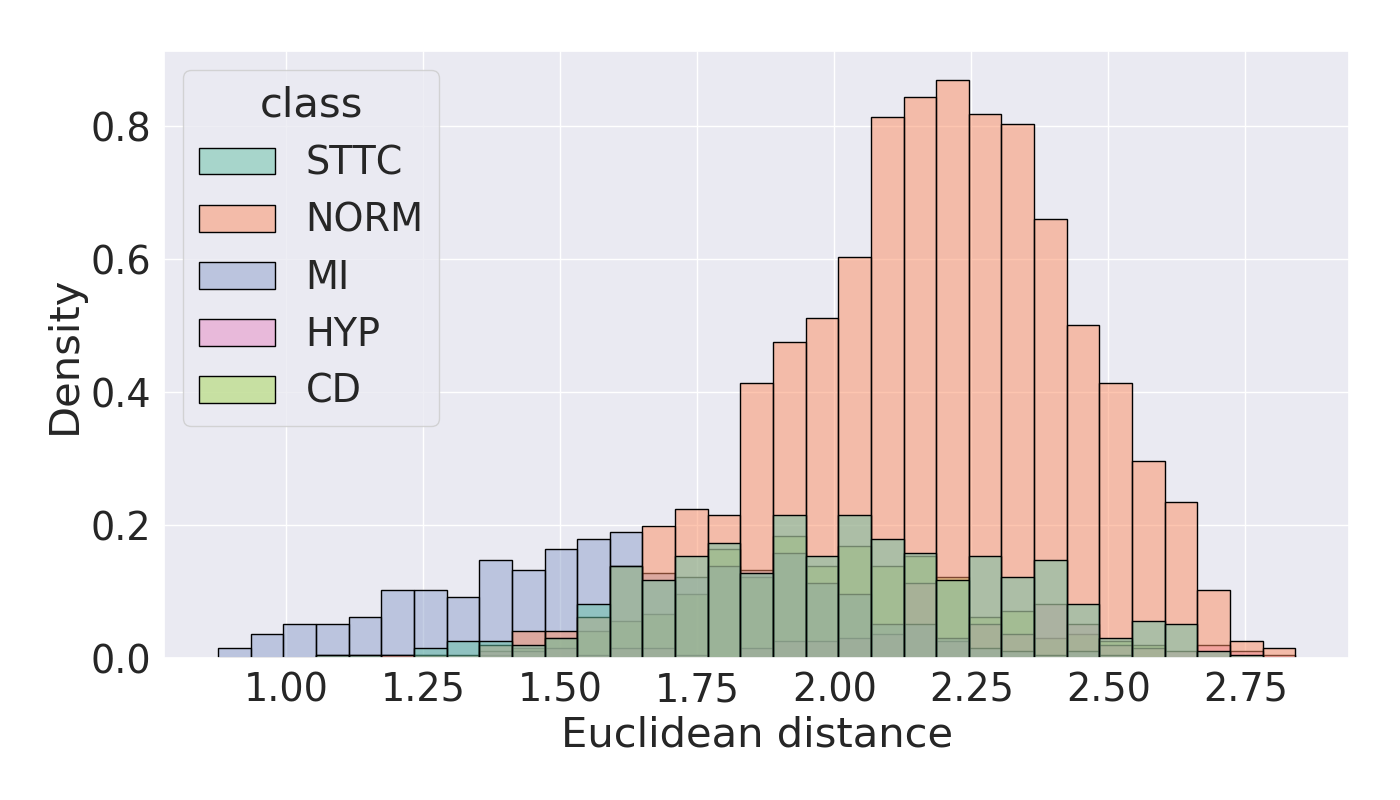}
	\end{subfigure}
	~
    \begin{subfigure}{0.48\textwidth}
    \centering
	\includegraphics[width=1\textwidth]{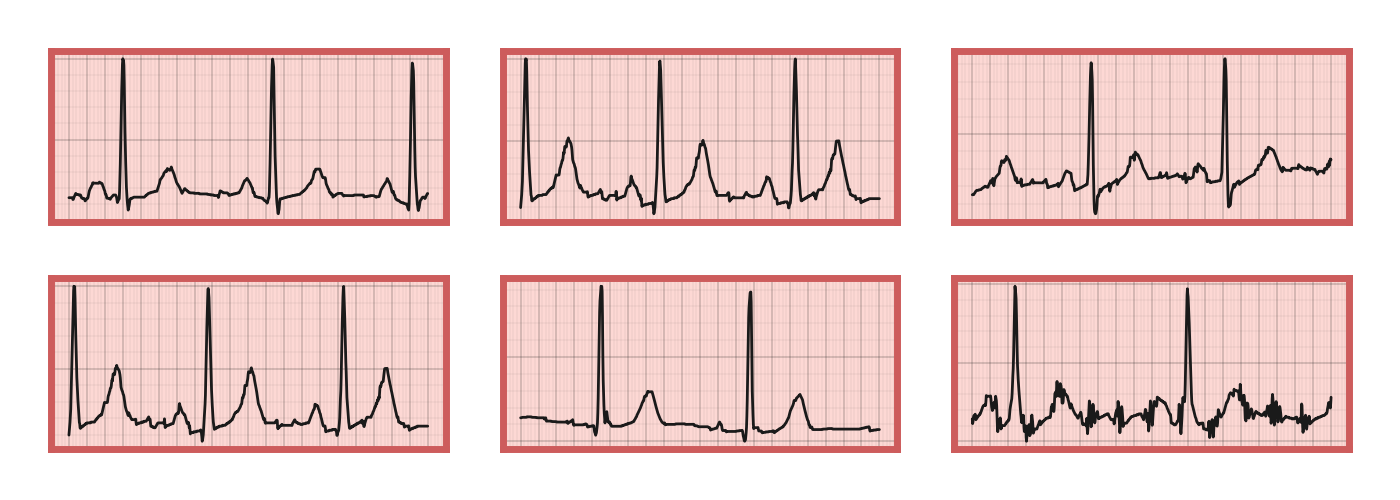}
	\caption{$\mathrm{DTC}$ query $\{\mathrm{MI},\mathrm{female},\mathrm{over \ 95}\}$}
	\label{fig:ptbxl_distances_cp}
	\end{subfigure}
	~
    \begin{subfigure}{0.48\textwidth}
    \centering
	\includegraphics[width=1\textwidth]{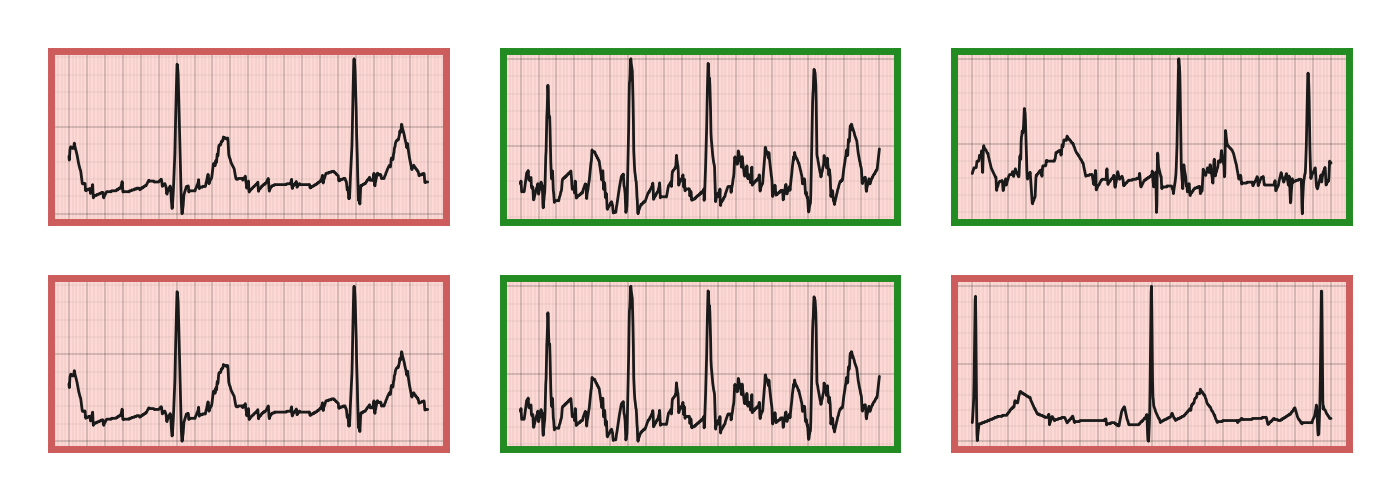}
	\caption{$\mathrm{CP \ CROCS}$ query $\{\mathrm{MI},\mathrm{female},\mathrm{over \ 95}\}$}
	\label{fig:ptbxl_distances_cp}
	\end{subfigure}
	~
    \begin{subfigure}{1\textwidth}
    \centering
	\includegraphics[width=0.48\textwidth]{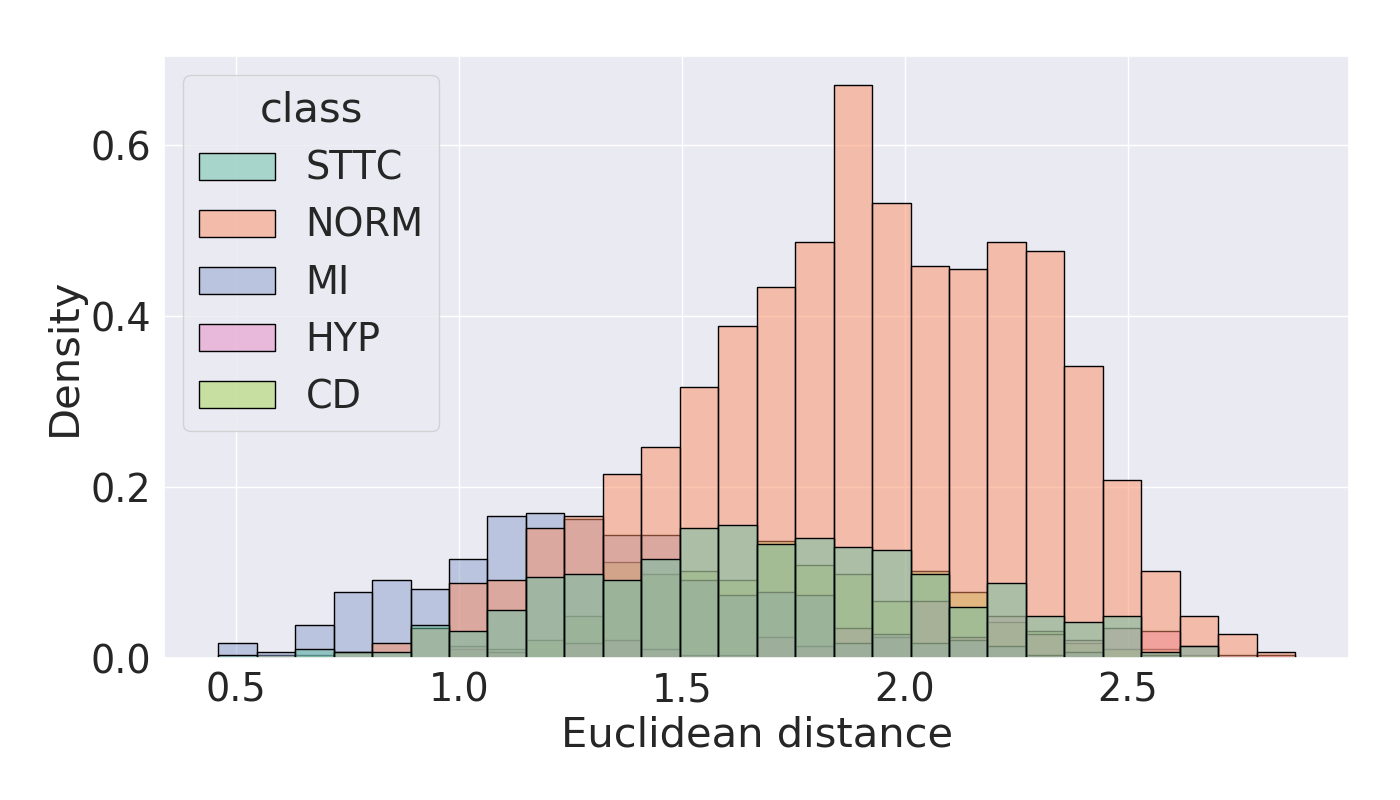}
	\end{subfigure}
	~
    \begin{subfigure}{1\textwidth}
    \centering
	\includegraphics[width=0.48\textwidth]{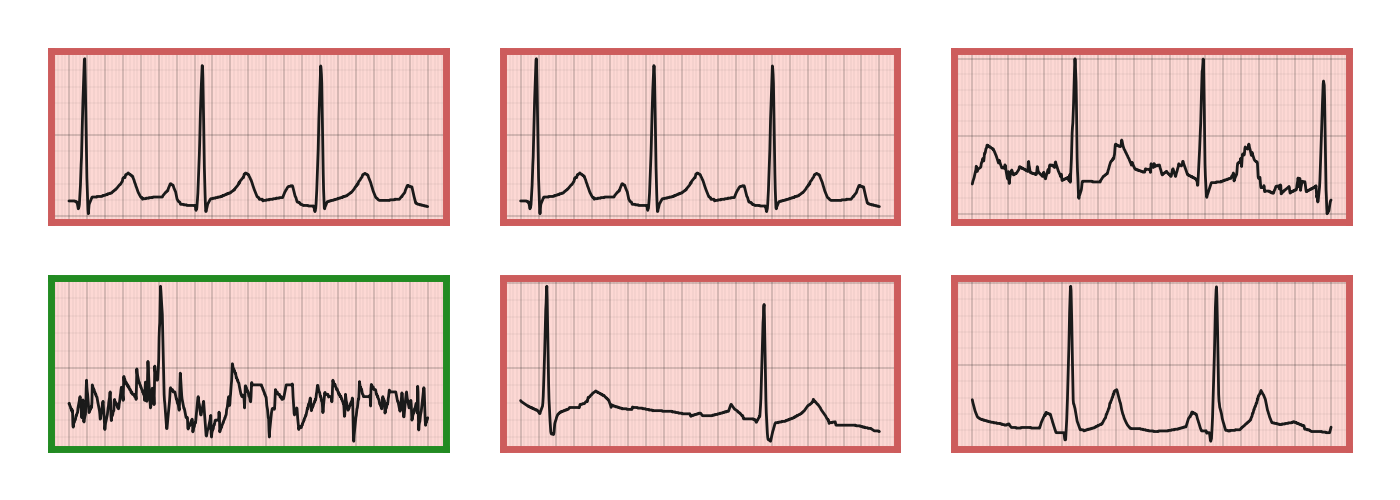}
	\caption{TP CROCS query $\{\mathrm{MI},\mathrm{female},\mathrm{over \ 95}\}$}
	\label{fig:ptbxl_distances_cp}
	\end{subfigure}
	\caption{\textbf{Qualitative retrieval performance of two distinct queries.} (\textbf{top row}) Euclidean distance from a query (a) $\mathrm{DTC}$ (b) $\mathrm{CP \ CROCS}$ or (c) $\mathrm{TP \ CROCS}$ query, to representations in the validation set of PTB-XL. (\textbf{bottom row}) Six closest cardiac signals to each query. Each query is associated with a set of patient attributes $\{ \mathrm{disease}, \mathrm{sex}, \mathrm{age} \}$. Retrieved cardiac signals with green borders indicate those whose class attribute matches that of the query. We show that the clinical prototype is closer to representations of the same class ($\mathrm{MI}$) and thus retrieve relevant cardiac signals.}
	\label{fig:ptbxl_top_k_ecgs}
\end{figure}

In this section, we continue our qualitative evaluation of the retrieval performance of various methods. In Fig.~\ref{fig:ptbxl_top_k_ecgs} (top row), we present the distributions of the Euclidean distance between the query (DTC-derived prototype or clinical prototype) and the representations in the validation set of PTB-XL. In Fig.~\ref{fig:ptbxl_top_k_ecgs} (bottom row), we illustrate the six cardiac signals that are closest to the respective prototypes. As with the results in the main manuscript, we find that the clinical prototype is closer to representations of the same class attribute than to those with a different class attribute. This is evident by the long tail of distance values exhibited between representation with $\mathrm{MI}$ and the clinical prototype $\{\mathrm{MI},\mathrm{female},\mathrm{over \ 95}\}$. This behaviour, which is non-existent for the DTC-derived prototype, can explain the relatively improved retrieval performance of clinical prototypes. This is further supported by the retrieved cardiac signals (Fig.~\ref{fig:ptbxl_top_k_ecgs} bottom row) where the DTC-derived prototype and the clinical prototype retrieve relevant instances $0\%$ and $50\%$ of the time, respectively.  

\clearpage

\section{Investigating marginal impact of design choices}
\label{appendix:marginal_impact}

In this section, we quantify the marginal impact of the design choices of our CROCS framework on the retrieval performance. In Table~\ref{table:marginal_retrieval_results}, we present the precision of retrieved cardiac signals when evaluated based on both partial and exact matches of attributes also represented by the query. Each query is a clinical prototype that is learned in a variant of the CROCS framework. These variants are shown in Sec.~\ref{sec:methods} of the main manuscript. We find that clinical prototypes learned via our full framework ($\mathcal{L}_{NCE-soft} + \mathcal{L}_{reg}$) add value relative to those learned under the $\mathrm{Hard}$ assignment framework. For example, at $K=1$, and when \# attribute matches $\geq2$, $\mathcal{L}_{NCE-hard}$, $\mathcal{L}_{NCE-soft}$ $\tau_{\omega} = \infty$, $\tau_{\omega} \neq \infty$, and $\mathcal{L}_{NCE-soft} + \mathcal{L}_{reg}$ achieve a precision of $27.5$, $51.5$, $58.5$, and $63.0$, respectively. 

\begin{table}[!h]
\scriptsize
\centering
\caption{\textbf{Marginal impact of design choices of CROCS on the precision of $\boldsymbol{K}$ retrieved representations, $\boldsymbol{v}$, in the validation set of Chapman and PTB-XL, that are closest to the query.} Results are shown for partial and exact matches of the attributes (\# attribute matches) represented by the query and retrieved cardiac signals, and are averaged across five random seeds. Brackets indicate standard deviation and bold reflects the top-performing method.}
    \label{table:marginal_retrieval_results}
    \centering
    \vskip 0.05in
    \begin{tabular}{c c | c c c }
        \toprule
        \# attribute & \multirow{2}{*}{Query} & \multicolumn{3}{c}{PTB-XL} \\
        matches & & $K=1$ & $5$ & $10$ \\
        \midrule
        
        \multirow{6}{*}{$\geq1$} & $\mathcal{L}_{NCE-hard}$ & $70.0$ {\scriptsize(3.9)} & $100.0$ {\scriptsize(0.0)} & $100.0$ {\scriptsize(0.0)} \\
        
        \cmidrule{2-5}
        & \multicolumn{4}{l}{$\mathcal{L}_{NCE-soft}$} \\
        \cmidrule{2-5}
        
        & $\tau_{\omega} = \infty$ & $91.5$ {\scriptsize(2.0)} & $100.0$ {\scriptsize(0.0)} & $100.0$ {\scriptsize(0.0)} \\
        
        & $\tau_{\omega} \neq \infty$ & $88.0$ {\scriptsize(6.0)} & $100.0$ {\scriptsize(0.0)} & $100.0$ {\scriptsize(0.0)} \\
        
        & $+ \mathcal{L}_{reg}$ & $\textbf{92.5}$ {\scriptsize(0.0)} & $100.0$ {\scriptsize(0.0)} & $100.0$ {\scriptsize(0.0)} \\

        \midrule 
        \multirow{6}{*}{$\geq2$} & $\mathcal{L}_{NCE-hard}$ & $27.5$ {\scriptsize(3.9)} & $67.5$ {\scriptsize(0.0)} & $93.0$ {\scriptsize(4.0)} \\
        
        \cmidrule{2-5}
        & \multicolumn{4}{l}{$\mathcal{L}_{NCE-soft}$} \\
        \cmidrule{2-5}
        
        & $\tau_{\omega} = \infty$ & $51.5$ {\scriptsize(2.0)} & $94.5$ {\scriptsize(1.0)} & $100.0$ {\scriptsize(0.0)} \\
        
        & $\tau_{\omega} \neq \infty$ & $58.5$ {\scriptsize(2.0)} & $100.0$ {\scriptsize(0.0)} & $100.0$ {\scriptsize(0.0)} \\
        
        & $+ \mathcal{L}_{reg}$ & $\textbf{63.0}$ {\scriptsize(1.9)} & $96.5$ {\scriptsize(2.0)} & $99.5$ {\scriptsize(1.0)} \\

        \midrule
        \multirow{6}{*}{$=3$} & $\mathcal{L}_{NCE-hard}$ & $7.0$ {\scriptsize(1.0)} & $16.0$ {\scriptsize(3.0)} & $26.5$ {\scriptsize(4.6)} \\

        \cmidrule{2-5}
        & \multicolumn{4}{l}{$\mathcal{L}_{NCE-soft}$} \\
        \cmidrule{2-5}

        & $\tau_{\omega} = \infty$ & $9.5$ {\scriptsize(1.0)} & $30.5$ {\scriptsize(4.0)} & $38.5$ {\scriptsize(3.0)} \\

        & $\tau_{\omega} \neq \infty$ & $12.5$ {\scriptsize(0.0)} & $36.5$ {\scriptsize(2.0)} & $43.5$ {\scriptsize(1.2)} \\
    
        & $+ \mathcal{L}_{reg}$ & $12.5$ {\scriptsize(0.0)} & $33.5$ {\scriptsize(3.0)} & $43.0$ {\scriptsize(4.0)} \\

        \bottomrule
\end{tabular}
\end{table}

\clearpage

\section{Discovering attribute-specific features within clinical prototypes}
\label{appendix:agglomerative_clustering}

\begin{figure}[!h]
    \centering
    \begin{subfigure}{0.8\textwidth}
    \centering
    \includegraphics[width=\textwidth]{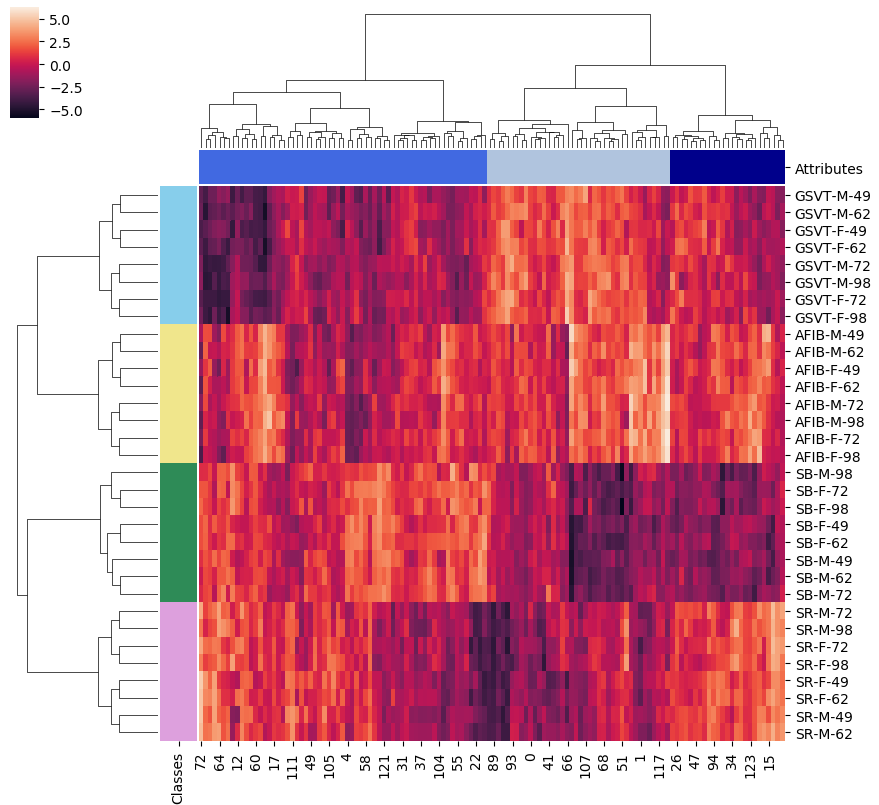}
    \end{subfigure}
    \caption{\textbf{Hierarchical agglomerative clustering of the clinical prototypes}. \textbf{(rows)} Clustering is performed using the 128-dimensional features resulting in 4 major clusters corresponding to the 4 classes. Clinical prototypes with similar attributes are also clustered together. The rows are labelled according to the attribute combination, $m$. \textbf{(columns)} Clustering is performed whereby each of the 128 \textit{features} is treated as an instance, resulting in 3 major clusters which are hypothesized to correspond to the 3 attributes: class, sex, and age. This suggests that disentangled, attribute-specific features may have been learned.}
    \label{fig:HAC_prototypes}
\end{figure}

We have shown that clinical prototypes can be deployed successfully for retrieval and clustering purposes while managing to capture relationships between attributes. In this section, we aim to quantify the relationship between clinical prototypes and explore their features further. In Fig.~\ref{fig:HAC_prototypes}, we illustrate a matrix of the clinical prototypes ($M=32$) along the rows and their corresponding features ($E=128$) along the columns. By implementing the hierarchical agglomerative clustering (HAC) algorithm, we cluster these clinical prototypes and arrive at the dendrogram presented along the rows of Fig.~\ref{fig:HAC_prototypes}. In addition to being correctly clustered according to class labels, they are also more similar to one another based on their attributes. This can be seen by the attribute combination descriptions in the right column. This finding supports our earlier claim that clinical prototypes do indeed capture relationships between attributes. 

Motivated by recent work on disentangled representations, whereby representations can be factorized into multiple sub-groups each of which correspond to a particular abstraction, we chose to cluster the \textit{features} of the clinical prototypes, resulting in the dendrogram presented along the columns of Fig.~\ref{fig:HAC_prototypes}. The intuition is that by clustering we may discover attribute-specific feature subsets. We show that these features can indeed be clustered into three main groups, potentially coinciding with our pre-defined attributes. Such a process can improve the interpretability of clinical prototypes and lead to insights about how they can be further manipulated for retrieval purposes, for instance, by altering a subset of features. 

\end{document}